\RequirePackage{lineno}
 
\documentclass[aps,prd,reprint,superscriptaddress,floats,floatfix]{revtex4-1}

\usepackage{subfig}
\usepackage{amsmath}

\usepackage{graphicx}
\usepackage{amssymb}

\usepackage{dcolumn}
\usepackage{bm}

\newcommand{\ttbar}{\ensuremath{t\bar{t}}}
\newcommand{\bbar}{\ensuremath{b\bar{b}}}

\newcommand{\ppbar}{\ensuremath{p\bar{p}}}

\newcommand{\qq}{\ensuremath{qq^{\prime}}}

\newcommand{\invfb}{\ensuremath{\rm fb^{-1}}}
\newcommand{\mbb}{\ensuremath{m_{bb}}}
\newcommand{\mqq}{\ensuremath{m_{qq}}}

\newcommand{\et}{\ensuremath{E_T}}

\newcommand{\dr}{\ensuremath{\Delta R}}

\newcommand{\gevcc}{\ensuremath{ \text{GeV/}c^{\text{2}}}}

\begin{document}



\preprint{Fermilab-PUB-XX-xx-X}

\title{Search for the Higgs boson in the all-hadronic final state using the CDF II detector}

\affiliation{Institute of Physics, Academia Sinica, Taipei, Taiwan 11529, Republic of China} 
\affiliation{Argonne National Laboratory, Argonne, Illinois 60439, USA} 
\affiliation{University of Athens, 157 71 Athens, Greece} 
\affiliation{Institut de Fisica d'Altes Energies, ICREA, Universitat Autonoma de Barcelona, E-08193, Bellaterra (Barcelona), Spain} 
\affiliation{Baylor University, Waco, Texas 76798, USA} 
\affiliation{Istituto Nazionale di Fisica Nucleare Bologna, $^z$University of Bologna, I-40127 Bologna, Italy} 
\affiliation{University of California, Davis, Davis, California 95616, USA} 
\affiliation{University of California, Los Angeles, Los Angeles, California 90024, USA} 
\affiliation{Instituto de Fisica de Cantabria, CSIC-University of Cantabria, 39005 Santander, Spain} 
\affiliation{Carnegie Mellon University, Pittsburgh, Pennsylvania 15213, USA} 
\affiliation{Enrico Fermi Institute, University of Chicago, Chicago, Illinois 60637, USA}
\affiliation{Comenius University, 842 48 Bratislava, Slovakia; Institute of Experimental Physics, 040 01 Kosice, Slovakia} 
\affiliation{Joint Institute for Nuclear Research, RU-141980 Dubna, Russia} 
\affiliation{Duke University, Durham, North Carolina 27708, USA} 
\affiliation{Fermi National Accelerator Laboratory, Batavia, Illinois 60510, USA} 
\affiliation{University of Florida, Gainesville, Florida 32611, USA} 
\affiliation{Laboratori Nazionali di Frascati, Istituto Nazionale di Fisica Nucleare, I-00044 Frascati, Italy} 
\affiliation{University of Geneva, CH-1211 Geneva 4, Switzerland} 
\affiliation{Glasgow University, Glasgow G12 8QQ, United Kingdom} 
\affiliation{Harvard University, Cambridge, Massachusetts 02138, USA} 
\affiliation{Division of High Energy Physics, Department of Physics, University of Helsinki and Helsinki Institute of Physics, FIN-00014, Helsinki, Finland} 
\affiliation{University of Illinois, Urbana, Illinois 61801, USA} 
\affiliation{The Johns Hopkins University, Baltimore, Maryland 21218, USA} 
\affiliation{Institut f\"{u}r Experimentelle Kernphysik, Karlsruhe Institute of Technology, D-76131 Karlsruhe, Germany} 
\affiliation{Center for High Energy Physics: Kyungpook National University, Daegu 702-701, Korea; Seoul National University, Seoul 151-742, Korea; Sungkyunkwan University, Suwon 440-746, Korea; Korea Institute of Science and Technology Information, Daejeon 305-806, Korea; Chonnam National University, Gwangju 500-757, Korea; Chonbuk National University, Jeonju 561-756, Korea} 
\affiliation{Ernest Orlando Lawrence Berkeley National Laboratory, Berkeley, California 94720, USA} 
\affiliation{University of Liverpool, Liverpool L69 7ZE, United Kingdom} 
\affiliation{University College London, London WC1E 6BT, United Kingdom} 
\affiliation{Centro de Investigaciones Energeticas Medioambientales y Tecnologicas, E-28040 Madrid, Spain} 
\affiliation{Massachusetts Institute of Technology, Cambridge, Massachusetts 02139, USA} 
\affiliation{Institute of Particle Physics: McGill University, Montr\'{e}al, Qu\'{e}bec, Canada H3A~2T8; Simon Fraser University, Burnaby, British Columbia, Canada V5A~1S6; University of Toronto, Toronto, Ontario, Canada M5S~1A7; and TRIUMF, Vancouver, British Columbia, Canada V6T~2A3} 
\affiliation{University of Michigan, Ann Arbor, Michigan 48109, USA} 
\affiliation{Michigan State University, East Lansing, Michigan 48824, USA}
\affiliation{Institution for Theoretical and Experimental Physics, ITEP, Moscow 117259, Russia}
\affiliation{University of New Mexico, Albuquerque, New Mexico 87131, USA} 
\affiliation{Northwestern University, Evanston, Illinois 60208, USA} 
\affiliation{The Ohio State University, Columbus, Ohio 43210, USA} 
\affiliation{Okayama University, Okayama 700-8530, Japan} 
\affiliation{Osaka City University, Osaka 588, Japan} 
\affiliation{University of Oxford, Oxford OX1 3RH, United Kingdom} 
\affiliation{Istituto Nazionale di Fisica Nucleare, Sezione di Padova-Trento, $^{aa}$University of Padova, I-35131 Padova, Italy} 
\affiliation{LPNHE, Universite Pierre et Marie Curie/IN2P3-CNRS, UMR7585, Paris, F-75252 France} 
\affiliation{University of Pennsylvania, Philadelphia, Pennsylvania 19104, USA}
\affiliation{Istituto Nazionale di Fisica Nucleare Pisa, $^{bb}$University of Pisa, $^{cc}$University of Siena and $^{dd}$Scuola Normale Superiore, I-56127 Pisa, Italy} 
\affiliation{University of Pittsburgh, Pittsburgh, Pennsylvania 15260, USA} 
\affiliation{Purdue University, West Lafayette, Indiana 47907, USA} 
\affiliation{University of Rochester, Rochester, New York 14627, USA} 
\affiliation{The Rockefeller University, New York, New York 10065, USA} 
\affiliation{Istituto Nazionale di Fisica Nucleare, Sezione di Roma 1, $^{ee}$Sapienza Universit\`{a} di Roma, I-00185 Roma, Italy} 

\affiliation{Rutgers University, Piscataway, New Jersey 08855, USA} 
\affiliation{Texas A\&M University, College Station, Texas 77843, USA} 
\affiliation{Istituto Nazionale di Fisica Nucleare Trieste/Udine, I-34100 Trieste, $^{ff}$University of Trieste/Udine, I-33100 Udine, Italy} 
\affiliation{University of Tsukuba, Tsukuba, Ibaraki 305, Japan} 
\affiliation{Tufts University, Medford, Massachusetts 02155, USA} 
\affiliation{University of Virginia, Charlottesville, VA  22906, USA}
\affiliation{Waseda University, Tokyo 169, Japan} 
\affiliation{Wayne State University, Detroit, Michigan 48201, USA} 
\affiliation{University of Wisconsin, Madison, Wisconsin 53706, USA} 
\affiliation{Yale University, New Haven, Connecticut 06520, USA} 
\author{T.~Aaltonen}
\affiliation{Division of High Energy Physics, Department of Physics, University of Helsinki and Helsinki Institute of Physics, FIN-00014, Helsinki, Finland}
\author{B.~\'{A}lvarez~Gonz\'{a}lez$^v$}
\affiliation{Instituto de Fisica de Cantabria, CSIC-University of Cantabria, 39005 Santander, Spain}
\author{S.~Amerio}
\affiliation{Istituto Nazionale di Fisica Nucleare, Sezione di Padova-Trento, $^{aa}$University of Padova, I-35131 Padova, Italy} 

\author{D.~Amidei}
\affiliation{University of Michigan, Ann Arbor, Michigan 48109, USA}
\author{A.~Anastassov}
\affiliation{Northwestern University, Evanston, Illinois 60208, USA}
\author{A.~Annovi}
\affiliation{Laboratori Nazionali di Frascati, Istituto Nazionale di Fisica Nucleare, I-00044 Frascati, Italy}
\author{J.~Antos}
\affiliation{Comenius University, 842 48 Bratislava, Slovakia; Institute of Experimental Physics, 040 01 Kosice, Slovakia}
\author{G.~Apollinari}
\affiliation{Fermi National Accelerator Laboratory, Batavia, Illinois 60510, USA}
\author{J.A.~Appel}
\affiliation{Fermi National Accelerator Laboratory, Batavia, Illinois 60510, USA}
\author{A.~Apresyan}
\affiliation{Purdue University, West Lafayette, Indiana 47907, USA}
% ...
\collaboration{CDF Collaboration\footnote{With visitors from $^a$University of Massachusetts Amherst, Amherst, Massachusetts 01003,
$^b$Istituto Nazionale di Fisica Nucleare, Sezione di Cagliari, 09042 Monserrato (Cagliari), Italy,
$^c$University of California Irvine, Irvine, CA  92697, 
$^d$University of California Santa Barbara, Santa Barbara, CA 93106
$^e$University of California Santa Cruz, Santa Cruz, CA  95064,
$^f$CERN,CH-1211 Geneva, Switzerland,
$^g$Cornell University, Ithaca, NY  14853, 
$^h$University of Cyprus, Nicosia CY-1678, Cyprus, 
$^i$University College Dublin, Dublin 4, Ireland,
$^j$University of Fukui, Fukui City, Fukui Prefecture, Japan 910-0017,
$^k$Universidad Iberoamericana, Mexico D.F., Mexico,
$^l$Iowa State University, Ames, IA  50011,
$^m$University of Iowa, Iowa City, IA  52242,
$^n$Kinki University, Higashi-Osaka City, Japan 577-8502,
$^o$Kansas State University, Manhattan, KS 66506,
$^p$University of Manchester, Manchester M13 9PL, England,
$^q$Queen Mary, University of London, London, E1 4NS, England,
$^r$Muons, Inc., Batavia, IL 60510,
$^s$Nagasaki Institute of Applied Science, Nagasaki, Japan, 
$^t$National Research Nuclear University, Moscow, Russia,
$^u$University of Notre Dame, Notre Dame, IN 46556,
$^v$Universidad de Oviedo, E-33007 Oviedo, Spain, 
$^w$Texas Tech University, Lubbock, TX  79609, 
$^x$Universidad Tecnica Federico Santa Maria, 110v Valparaiso, Chile,
$^y$Yarmouk University, Irbid 211-63, Jordan,
$^{gg}$On leave from J.~Stefan Institute, Ljubljana, Slovenia, 
}}
\noaffiliation


\date{\today}


\begin{abstract}

We report on a search for the production of the Higgs boson decaying to two bottom quarks accompanied by two additional quarks. The data sample used corresponds to an integrated luminosity of approximately 4\,\invfb\, of \ppbar\, collisions at $\sqrt{s}=1.96$\,TeV recorded by the CDF II experiment. This search includes twice the integrated luminosity of the previous published result, uses analysis techniques to distinguish jets originating from light flavor quarks  and those from gluon radiation, and adds sensitivity to a Higgs boson produced by vector boson fusion.  We find no evidence of the Higgs boson and place limits on the Higgs boson production cross section for Higgs boson masses between 100\,\gevcc\, and 150\,\gevcc\, at the 95\% confidence level.  For a Higgs boson mass of 120\,\gevcc\, the observed (expected) limit is 10.5\,(20.0) times the predicted Standard Model cross section. 
\end{abstract}

\pacs{Valid PACS appear here}

\maketitle

\section{Introduction}

The Higgs boson remains the only undiscovered particle of the standard model (SM) of particle physics.  It is the physical manifestation of the mechanism which provides mass to fundamental particles~\cite{Higgs:1964ia,Englert1964Broken-Symmetry}. Direct searches at the LEP collider have excluded a Higgs boson mass $m_{H} < 114.4$\,\gevcc\, at 95\% confidence level (CL)~\cite{Barate:2003sz}, while the Tevatron collaborations have excluded a Higgs boson mass between 163\,\gevcc\, and 166\,\gevcc\, at 95\% CL~\cite{PhysRevLett.104.061802}.  The Tevatron collaborations have reported a preliminary update which extends the exclusion region for a Higgs boson mass between 158 and 173\,\gevcc~\cite{Collaboration2010Combined-CDF-an}.  Global fits to precision electroweak measurements set a one-sided 95\% CL upper limit on $m_{H}$ at 157\,\gevcc~\cite{Alcaraz:2009jr}.  


This article presents the results of a search for the Higgs boson using an integrated luminosity of  4\,\invfb\, of \ppbar\, collision data at $\sqrt{s} = 1.96$\,TeV recorded by the Collider Detector at Fermilab (CDF II).  We search for a Higgs boson decaying to a pair of bottom-quark jets (\bbar) accompanied by two additional quark jets (\qq) for Higgs mass $100 \leq m_{H} \leq 150$\,\gevcc.  This search is most sensitive to a Higgs boson with low mass, $m_{H} < 135\,\gevcc$, where the Higgs boson decay to \bbar\, is dominant~\cite{Djouadi:1997yw}. The two production channels studied are associated production and vector boson fusion (VBF).  The associated production channel is $\ppbar \rightarrow VH \rightarrow \qq \, \bbar$, where $V$ is a $W/Z$ vector boson, which decays to a pair of quarks. The hadronic branching fraction of $V$  to \qq\, is $\simeq 70\%$~\cite{0954-3899-37-7A-075021}.  In the VBF channel, $\ppbar \rightarrow \qq H \rightarrow \qq \, \bbar$, the incoming partons each radiate a vector boson and the two vector bosons fuse to form a Higgs boson.  

Low-mass Higgs boson searches at CDF have concentrated on signatures that are a combination of jets, leptons and missing transverse energy which help to reduce the backgrounds but the signal yields are small~\cite{Aaltonen2009Search-for-the-, PhysRevLett.103.101802, PhysRevLett.104.141801}.  The hadronic modes used in this search exploit the larger branching fraction and thus have the largest signal yields among all the search channels at CDF.  The major challenge for this search is the modeling and suppression of the large background from QCD multijets (referred to as QCD for brevity).



A previous letter on the search for the Higgs boson in the all-hadronic channel was published using an integrated luminosity of 2\,\invfb\,~\cite{PhysRevLett.103.221801}.  This article has lowered the expected limit by a factor of two: a factor of $\approx\sqrt{2}$ from doubling the analyzed data and a factor of 1.4 from improvements to the analysis which are discussed in this article.

\section{The Tevatron and the CDF II Detector}

The CDF II detector, designed to study \ppbar\, collisions, is both an azimuthally and forward-backward symmetric. 
It is described in detail in Refs.~\cite{Acosta:2004yw,Acosta:2004hw,Abulencia:2005ix} and references therein. CDF II uses a cylindrical coordinate system in which the $z$ axis aligned along the proton beam direction,  $\theta$ is the polar angle relative to the $z$-axis and $\phi$ is the azimuthal angle relative to the $x$-axis. The pseudorapidity is defined as $\eta \equiv -\ln(\tan \theta/2)$.  The transverse energy is $\et \equiv E \sin \theta$. Jets are defined by  a cluster of energy in the calorimeter deposited inside a cone of radius $\dr = \sqrt{(\Delta\phi)^{2} + (\Delta\eta)^{2} }  = 0.4$  as reconstructed by the {\sc JetClu} algorithm ~\cite{Abe:1991ui}.  Corrections are applied to the measured jet energy to account for detector calibrations, multiple interactions, underlying event and energy outside of the jet cone~\cite{Bhatti:2005ai}.  

The data for this search were collected by two multijet triggers.  The first 2.8\,\invfb\, used a trigger which selected  at least four jet clusters with $\et \ge 15\,$GeV for each jet and a total $\et \ge 175$\,GeV.  This trigger was used in the previous result~\cite{PhysRevLett.103.221801}.  The remaining 1.1\,\invfb\, were recorded with a new trigger which selected at least three jet clusters with $\et \ge 20$\,GeV for each jet and a total $\et \ge 130$\,GeV. 
The new trigger improved the acceptance for a low-mass Higgs boson by 45\% at Higgs mass of 100\,\gevcc\, and by 
20\% at Higgs mass of 150\,\gevcc. The improvement was mainly due to lowering the total \et\, criteria in the new trigger. However the gain in the signal acceptance of the new trigger was diminished after the event selection criteria which are described in the next section.



\section{Event Selection}
Events with isolated leptons or missing transverse energy significance~\footnote{missing transverse energy significance is defined as  the ratio of the total missing transverse energy to the square root of the total transverse energy}  $>$ 6 are removed to avoid any overlap with other low-mass Higgs analyses at CDF II.
%
%
The data are refined further by selecting events with four or five jets where each jet has \et\, $>$ 15\,GeV and $|\eta| < 2.4$. The selected jets are ordered by descending jet-\et\, and any fifth jet plays no further role. The scalar sum of the four leading jets' \et\, is required to be $> 220$\,GeV, and exactly two of the four leading jets are required to be identified (``tagged") as bottom-quark jets ($b$ jet).   The scalar sum \et\, cut reduces the contribution of the QCD background.  A $b$ jet is identified by its  displaced vertex, as defined by the {\sc SecVtx} algorithm~\cite{Acosta:2004hw}, or by using the probability that the tracks within the jet  are inconsistent with originating from the primary \ppbar\, collision as defined by the {\sc JetProb} algorithm~\cite{Abulencia:2006kv}.  The final four jets are labeled as $b_1$, $b_2$, $q_1$, $q_2$ where $b$($q$) are tagged (untagged) jets and $\et^{b_1,q_1}  > \et^{b_2,q_2}$.









The signal/background ratio is enhanced by dividing the data into two non-overlapping $b$-tagging categories:  SS when both jets are tagged by {\sc SecVtx}, SJ when one jet is tagged by {\sc SecVtx} and the other by {\sc JetProb}.   For a jet tagged by both algorithms, {\sc SecVtx} takes precedence as it has a lower rate of misidentifying a light flavor jet as a $b$ jet.   The previous 2\,\invfb\, search only included the SS category~\cite{PhysRevLett.103.221801} and  the addition of the SJ category increases the signal acceptance by 36\%.  Other $b$-tagging combinations, such as both $b$ jets selected by {\sc JetProb},  were not considered in this search as the relative increase in the background is much larger than that for the signal.



\begin{figure}[!]
\begin{center}
\hspace{-1cm}
\includegraphics[width=6.5cm]{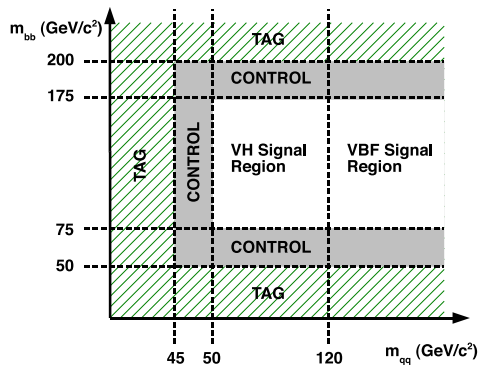}
\caption{$\mbb-\mqq$ plane:  This plane illustrates the {\sl VH} and VBF signal regions used to select {\sl VH} and VBF candidates.  The {\sc Tag} region is used to derive a model of the QCD background. The  {\sc Control} region measures the systematic uncertainty of the QCD model.}
\label{FIG:MbbMqqPlane}
\end{center}
\end{figure}

The data are divided into  {\sl VH} and VBF candidates defined by the invariant masses of the $b_1 b_2$  pair, \mbb, and the $q_1 q_2$  pair, \mqq. {\sl VH} candidates have $75 < \mbb < 175$\,\gevcc\, and $50 < \mqq < 120$\,\gevcc . VBF candidates have  $75 < \mbb < 175$\,\gevcc\, and  $\mqq > 120$\,\gevcc. The typical \mbb\, dijet mass resolution is $\sim 18\%$~\cite{PhysRevLett.104.141801}. These {\sl VH} and VBF signal regions are illustrated in Fig.~\ref{FIG:MbbMqqPlane}. We search for  Higgs bosons produced via {\sl VH} and VBF exclusively in the {\sl VH} and VBF signal regions, respectively. The division of events is based on the different kinematics of the two processes. The {\sl VH} channel has two mass resonances:  \mbb\, from the Higgs boson decay and \mqq\, from the $V$ decay.  The VBF channel  shares the same \mbb\, Higgs boson mass resonance but there is no accompanying resonance for \mqq.  The $q$ jets in VBF tend to have a large $\eta$ separation which results in larger values of \mqq.  The cut of $\mqq > 120$\,\gevcc\, optimizes the VBF signal over background ratio.  The acceptance for {\sl VH} and VBF events varies from 2\% to 3\%  for $100\,\gevcc < m_{H} < 150\,\gevcc$.  As {\sl VH} and VBF candidates are also split by the two $b$-tagging categories, there are 4 independent samples (channels) which are studied:  {\sl VH}-SS; {\sl VH}-SJ; VBF-SS; VBF-SJ.

\section{Signal and Background Samples}

The data are compared to a model of the signal and background composed of QCD, \ttbar,  $Z (\rightarrow \bbar/c\bar{c})$ + Jets ($Z$+jets), single-top, $W + \bbar/c\bar{c}$  ($W$+HF), and $WW/WZ/ZZ$ (diboson) events.  The signal and non-QCD backgrounds are modeled by Monte Carlo (MC) simulation. The {\sl VH} and VBF production are generated by {\sc pythia}~\cite{Sjostrand:2000wi}, combined with a {\sc geant}-based~\cite{GEANT} simulation of the CDF II detector~\cite{2003physics...6031G}. The non-QCD MC is described in detail in Ref.~\cite{PhysRevLett.103.221801} and normalized to next-to-leading order cross sections.  All the MC samples include the trigger simulation and their trigger efficiencies are corrected as described in Ref.~\cite{PhysRevLett.103.221801}. The  QCD background shape is modeled by a data-driven technique developed in Ref.~\cite{PhysRevLett.103.221801} and described in detail below. The expected signal yields of the four channels are 7.8({\sl VH}-SS), 2.9({\sl VH}-SJ), 3.2(VBF-SS), and 1.2(VBF-SJ) for $m_{H} = 120\,\gevcc$. The total backgrounds are about 17000({\sl VH}-SS), 9300({\sl VH}-SJ), 18000(VBF-SS), and 9500(VBF-SJ). The background composition is  $\sim$98\% QCD (Table~\ref{Table:BackgroundSources}).

\begingroup
\squeezetable
\begin{table}
\caption{Expected number of non-QCD background and  {\sl VH}/VBF signal with observed number of events for the four channels.  Statistical and systematic uncertainties are combined in quadrature where systematic uncertainties dominate.  The number of {\sl VH}(VBF) events are exclusive to the {\sl VH}(VBF) channels.  The difference between data and non-QCD are assumed to be QCD.}
\begin{ruledtabular}
\begin{tabular}{l c c c c} 
                                                        & {\sl VH}-SS                 & {\sl VH}-SJ                 & VBF-SS                     & VBF-SJ \\ \hline
\ttbar                                               & $281.7 \pm 45.6$  & $115.3 \pm 19.9$  & $177.3 \pm 28.7$   & $75.7 \pm 13.1$ \\
Single-top                                     & $44.1 \pm 7.1$       & $17.7 \pm 3.1$       &  $17.2 \pm 2.8$       & $10.0 \pm 1.7$ \\
$Z $ + Jets                                    & $127.5 \pm 65.8$  & $55.4 \pm 28.8$     &  $135.0 \pm 69.7$  & $62.9 \pm 32.7$ \\
$W$ + HF                                      & $27.9 \pm 14.4$    & $12.0 \pm 6.2$       &   $4.8 \pm 2.5$        & $3.3 \pm 1.7$ \\
Diboson                                        & $11.4 \pm 1.6$       & $8.5 \pm 1.3$         &  $5.3 \pm 0.7$          & $3.8 \pm 0.6$  \\  \hline
Total                                              & $492.6 \pm 81.7$   & $208.9 \pm 35.7$  & $339.6 \pm 75.5$   & $155.7 \pm 35.3$ \\
\multicolumn{5}{l}{Non-QCD}                                                 \\  \hline \hline
\multicolumn{5}{l}{Higgs Signal ($m_{H} = 120\,\gevcc$)} \\ 
{\sl VH}                                                & $7.8 \pm 1.0$         & $2.9 \pm 0.4$         &                                    &             \\ 
VBF                                               &                                   &                                  &  $3.2 \pm 0.4$          & $1.2 \pm 0.2$  \\ \hline
Data                                              & 16857                      & 9341                        &  17776                       & 9518  \\  
\end{tabular}
\end{ruledtabular}
\label{Table:BackgroundSources}
\end{table}
\endgroup

\section{QCD Modeling}


The shape of the dominant QCD background is modeled using a data-driven method known as the tag rate function (TRF) and is described in detail in~\cite{PhysRevLett.103.221801}. The TRF is applied to a QCD dominated data sample of events with at least one {\sc SecVtx} $b$-tagged jet (single-tagged events) to predict the distribution of events with exactly two $b$-tagged jets (double-tagged events). For each single-tagged event, the TRF gives the probability of each additional jet, called a {\em probe} jet, to be a second $b$-tagged jet.  The TRF is parameterized  as a function of three variables:  the \et\, and $\eta$ of the probe jet and $\dr$ between the tagged $b$ jet and probe jet.  The choice of variables used to parameterize the TRF is motivated by the kinematics of the QCD background and the characteristics of the $b$-tagging algorithms.  As the behavior of the {\sc SecVtx} and {\sc JetProb} $b$-tagging algorithms are not identical,  there is a TRF for SS and another TRF for SJ. The TRF is measured using jets in the {\sc Tag} region (Fig.~\ref{FIG:MbbMqqPlane}), defined as $\mqq < 45$\,\gevcc, $\mbb < 50$\,\gevcc\, and $\mbb > 200$\,\gevcc\, which is not in the {\sl VH} and VBF signal regions.

\section{Jet Moment}
\label{SEC:JetMoment}

The {\sl VH} and VBF $q$-jets are mostly quark jets while QCD $q$-jets are a mixture of gluon and quark jets. As gluon jets, on average,  tend to be broader than quark jets,  any variable related to the jet width is an additional tool to discriminate the Higgs signal from QCD.
  
 In this article, we use the jet $\phi$($\eta$) moment, $\langle \phi \rangle$ ($\langle \eta \rangle$)~\cite{Aaltonen2010Measurement-of-},  of $q$ jets which measures the jet width along the $\phi$($\eta$) axis. The jet $\phi$ and $\eta$ moments are defined by 
\begin{subequations}
\begin{equation}
\langle \phi \rangle = \sqrt{ 
{\Large \sum_{\text{towers}}}  \left[  
\frac{ \et^{\text{tower}} }{ \et^{\text{jet}}  } \;
\Bigl( 
\Delta\phi( \phi_{\text{tower}} , \phi_{\text{jet}} )
\Bigr)^{2} 
\right]  
}
\label{EQN:JetPhiMomentDefinition}
\end{equation}
\begin{equation}
\langle \eta \rangle = \sqrt{ 
{\Large \sum_{\text{towers}}}  \left[  
\frac{ \et^{\text{tower}} }{ \et^{\text{jet}}  } \;
\Bigl( 
\eta_{\text{tower}} - \eta_{\text{jet}} 
\Bigr)^{2} 
\right]  
}
\end{equation}
\label{EQN:JetMomentDefinition}
\end{subequations}
where the $\phi$ and $\eta$ jet moments are summed over the calorimeter towers forming the jet and depends on the tower-\et\, ($\et^{\text{tower}}$), the jet-\et\, ($\et^{\text{jet}}$),  the tower's $\phi$($\eta$) position, $\phi_{\text{tower}}$
($\eta_{\text{tower}}$), and the jet's $\phi$($\eta$) position, $\phi_{\text{jet}}$($\eta_{\text{jet}}$). The function 
$\Delta\phi( \phi_{\text{tower}} , \phi_{\text{jet}} )$  in Eq.~\ref{EQN:JetPhiMomentDefinition} is the smallest angular difference between $\phi_{\text{tower}}$ and $\phi_{\text{jet}}$.  The jet moment is a measure of the jet's width.




We checked whether the MC simulation of the quark jet moment matches the data. Gluon jet moments were not 
checked as the Higgs $q$-jet,  modelled by MC, are mostly quark jets  whereas gluon jets only appear in QCD which 
is derived from data. The hadronic $W$ decay from  $t\bar{t} \rightarrow bW \bar{b}W \rightarrow bl\nu + 
\bar{b}qq^{\prime}$, where $l$ is an electron or a muon, provides a source of quark jets.  The event selection from Ref.~\cite{Collaboration2010First-Measureme} was used to extract a \ttbar\, data sample which is 86\% \ttbar. The complete sample composition is described in Ref.~\cite{Collaboration2010First-Measureme}. The leading  untagged jet pair whose invariant mass is  $80 \pm 30$\,\gevcc\, is assumed to be the quark jets from the hadronic $W$ boson decay. The same event and leading untagged jet pair selection is applied to \ttbar\, MC to compare with data. 


The jet moment depends not only on the parton initiating the jet but also on the $\et^{\rm jet}$, $\eta_{\rm jet}$,
and the number of primary vertices in the event ($N_{\rm Vtx}$) which are not guaranteed to be the same for data and MC.  The dependencies are removed by rescaling the measured jet moment to a common reference of  $\et^{\rm jet}$=50\,\gevcc, $\eta_{\rm jet}$=0 and $N_{\rm Vtx}$=1, as measured in data. The rescaling for $\langle \phi \rangle$ is performed using
%
%
%
\begin{subequations}
\begin{multline}
  \langle \phi \rangle^{\prime}_{Data}  =  \langle \phi \rangle_{Data}  \qquad \times  \\
  \frac{f_{Data}^{\phi}(\et^{\rm jet}=50\,\gevcc,\eta_{\rm jet}=0,N_{\rm Vtx}=1)}{ f_{Data}^{\phi} (\et^{\rm jet},\eta_{\rm jet},N_{\rm Vtx})}
\end{multline}
\begin{multline}
  \langle \phi \rangle^{\prime}_{MC} =  \langle \phi \rangle_{MC} \qquad \times  \\  
  \frac{f_{Data}^{\phi} (\et^{\rm jet} =50\,\gevcc,\eta_{\rm jet} =0,N_{\rm Vtx}=1)}{  f_{MC}^{\phi} (\et^{\rm jet} ,\eta_{\rm jet},N_{\rm Vtx})}
\end{multline}
\label{EQN:RescaleJetMoment}
\end{subequations}                                                                     
where $f_{Data}^{\phi}(\et^{\rm jet},\eta_{\rm jet},N_{\rm Vtx})$ and $f_{MC}^{\phi}(\et^{\rm jet},\eta_{\rm jet},N_{\rm Vtx})$ are the 
$\langle \phi \rangle$ parameterizations for data and MC, respectively. 
$\langle \phi \rangle_{Data}$($\langle \phi \rangle_{MC}$) are the measured $\langle \phi \rangle$ for data(MC), and 
$\langle \phi \rangle_{Data}^{\prime}$($\langle \phi \rangle_{MC}^{\prime}$) are the rescaled values.   $\langle \eta \rangle$ are rescaled in a similar way but has a separate $\langle \eta \rangle  $ parameterization for data and MC.

\begin{figure*}[!]
\subfloat[\ttbar\, $\eta$ moment] {
\includegraphics[width=8cm]{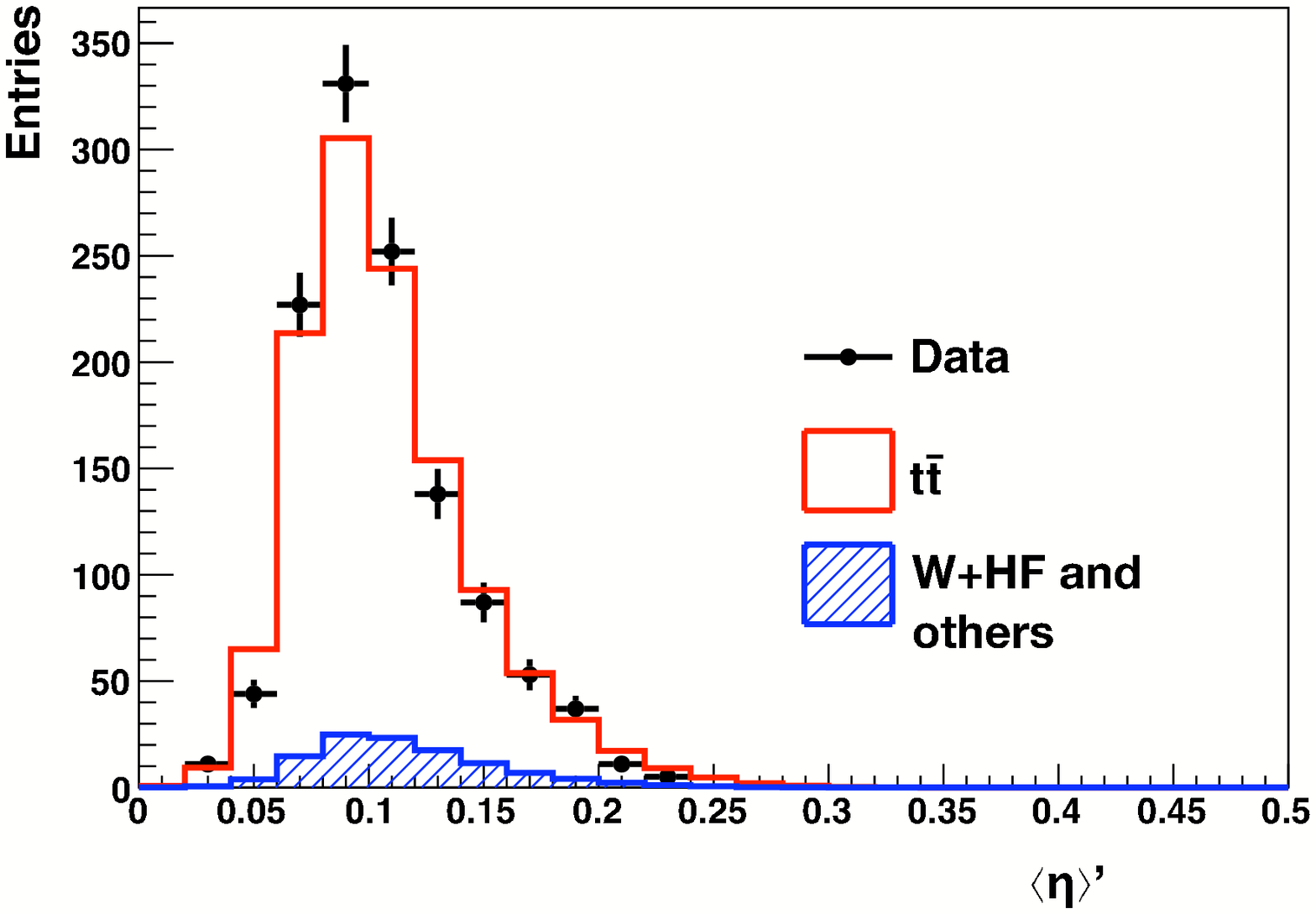}
}
%
\subfloat[\ttbar\, $\phi$ moment] {
\includegraphics[width=8cm]{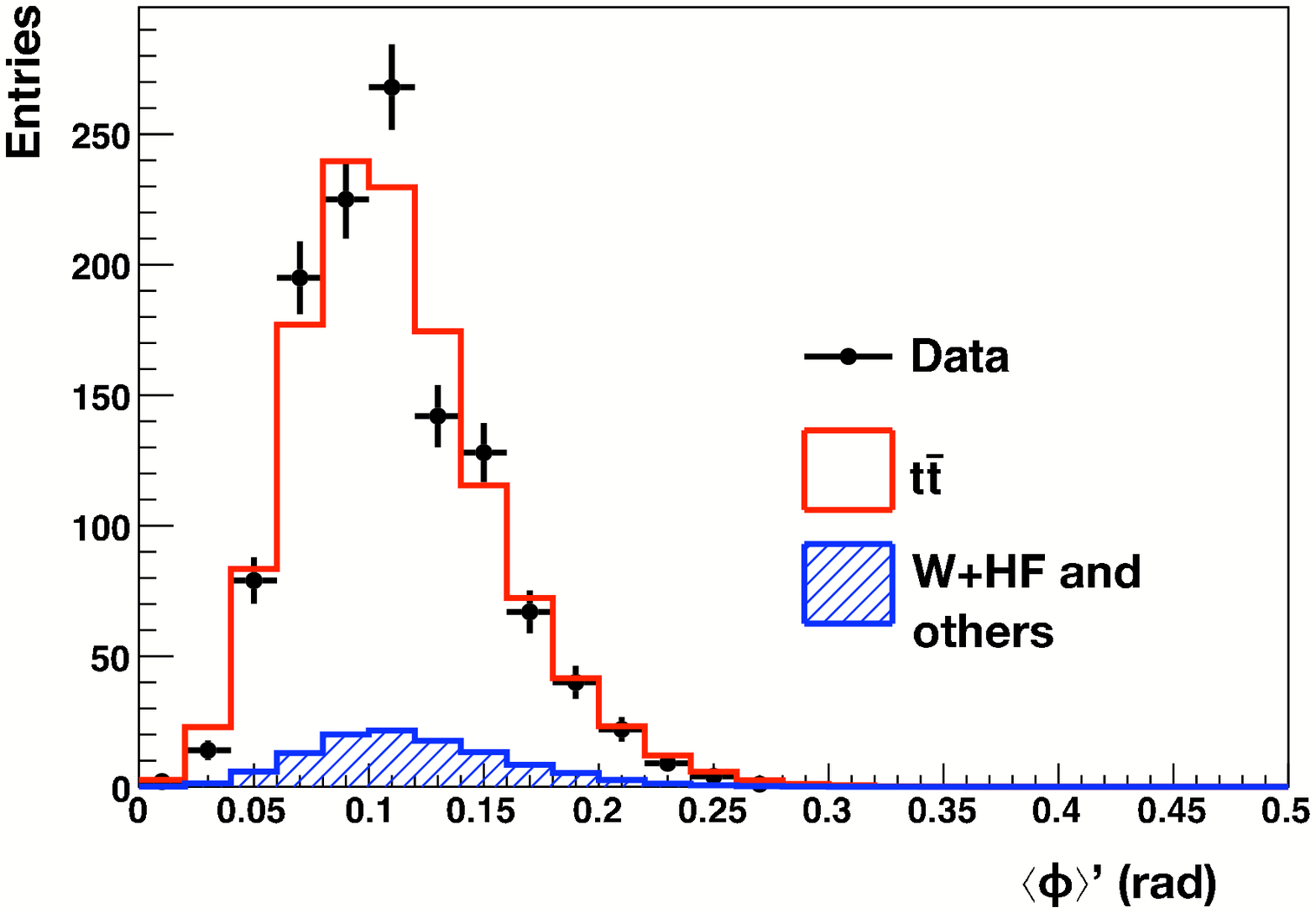}
}
\caption{Comparisons of the jet moments of \ttbar\, data  to \ttbar\, and $W$+HF, $Z$+jets, single top, diboson, misidentified $b$-jets ($W$+HF and others) MC. Only after applying all corrections does the MC agree with data.}
\label{FIG:ttbarJetMoments}
\end{figure*}

 
After the measured jet moments are rescaled, the MC required an additional shift in $\phi$ and $\eta$ of $\sim\!+2\%$ 
to agree with the data. Half of this offset was used as an estimate of the systematic uncertainty of the MC jet moment.  Figure~\ref{FIG:ttbarJetMoments} compares the jet moments of data to the simulated \ttbar\, signal and background MC which are in the same fractions as measured in data.  Only after applying all corrections does the MC agree with the data.


As an additional check,   the $\langle \phi \rangle^{\prime}_{MC}$ and $\langle \eta \rangle^{\prime}_{MC}$  of \ttbar\, MC was compared with {\sl VH} and VBF MC. The average jet moments of the MC samples were expected to
be identical to the $q$-jets from \ttbar\, as the Higgs signal are just quark jets. The jet moments from the {\sl VH} sample agreed with \ttbar. However there was a  disagreement of 5\% between VBF MC and \ttbar\, MC for jets with $|\eta| > 1.1$.  Half of this difference was used as an additional systematic uncertainty for the VBF jet moment.

\section{Neural Network}


The large background precludes the use of simple variables, such as \mbb,  to search for a Higgs boson signal.  An artificial neural network (NN), from the {\sc tmva} package~\cite{HoeckerTMVA---Toolkit-}, is trained to search exclusively for {\sl VH}(VBF) Higgs bosons in the {\sl VH}(VBF) signal region. The NN is trained using {\sl VH}(VBF) signal and TRF QCD prediction as background. 
As the kinematics for {\sl VH} and VBF Higgs signals are different, a dedicated NN for each signal is trained. The NN training variables for the {\sl VH}  NN are \mbb, \mqq, 
rescaled $q_1$  $\phi$ moment $\langle \phi_{1} \rangle^{\prime}$,  
rescaled $q_1$  $\eta$ moment $\langle \eta_{1} \rangle^{\prime}$,  
rescaled $q_2$  $\phi$ moment $\langle \phi_{2} \rangle^{\prime}$,
rescaled $q_2$  $\eta$ moment $\langle \eta_{2} \rangle^{\prime}$,
%
%
%
%
%
the cosine of the helicity angle $\cos \theta^{*}_{q_{1}}$~\footnote{$\cos \theta^{*}_ {q_{1}}$ is the cosine helicity angle of $q_1$. The $q_1$ helicity angle, $ \theta^{*}_{q_{1}}$,  is defined to be the angle between the momentum of $q_1$ in the $q_1-q_2$ rest frame and the total momentum of $q_1-q_2$ in the lab frame.}, the cosine of the leading jet scattering angle in the four jet rest frame, $\cos \theta_3$~\footnote{$\cos{\theta_{3}}$  is defined in a three jet rest frame as the cosine of the leading jet scattering angle. We reduce from four jets to three jets by combining the two jets with the lowest dijet mass.  Thus $\cos{\theta_{3}} = \frac{\vec{P_{AV}} \cdot \vec{P_{3}}}{|\vec{P_{AV}}|\vec{P_{3}}|}$, where $\vec{P_{3}}$ is the third jet and $\vec{P_{AV}}$ is the vector sum of the three jets in the lab frame~\cite{Geer:1995mp}.}, and  $\chi$ which is a measure of whether both the $b$-jet pair and $q$-jet pair are from a Higgs boson and $V$ decay, respectively.  $\chi$ is defined as the minimum of $\chi_{W}$ and $\chi_{Z}$ where $\chi_{W}$ is defined as  $\chi_{W}   =  \sqrt{(M_{W} - M_{qq})^2 + (M_H - M_{bb})^2}$ and a similar expression exists for $\chi_{Z}$. 
For the VBF channel, the neural net inputs are \mbb, \mqq, $\langle \phi_{1} \rangle^{\prime}$, $\langle \eta_{1} \rangle^{\prime}$, $\langle \phi_{2} \rangle^{\prime}$, and $\langle \eta_{2} \rangle^{\prime}$.



The two $b$-tagging categories have similar kinematic distributions which allows the same NN to be used for SS and SJ events. The NN is trained with SS events as it has the better signal/background ratio.

\begin{figure*}[p!]
%
\subfloat[] {
\includegraphics[width=6cm, angle=90]{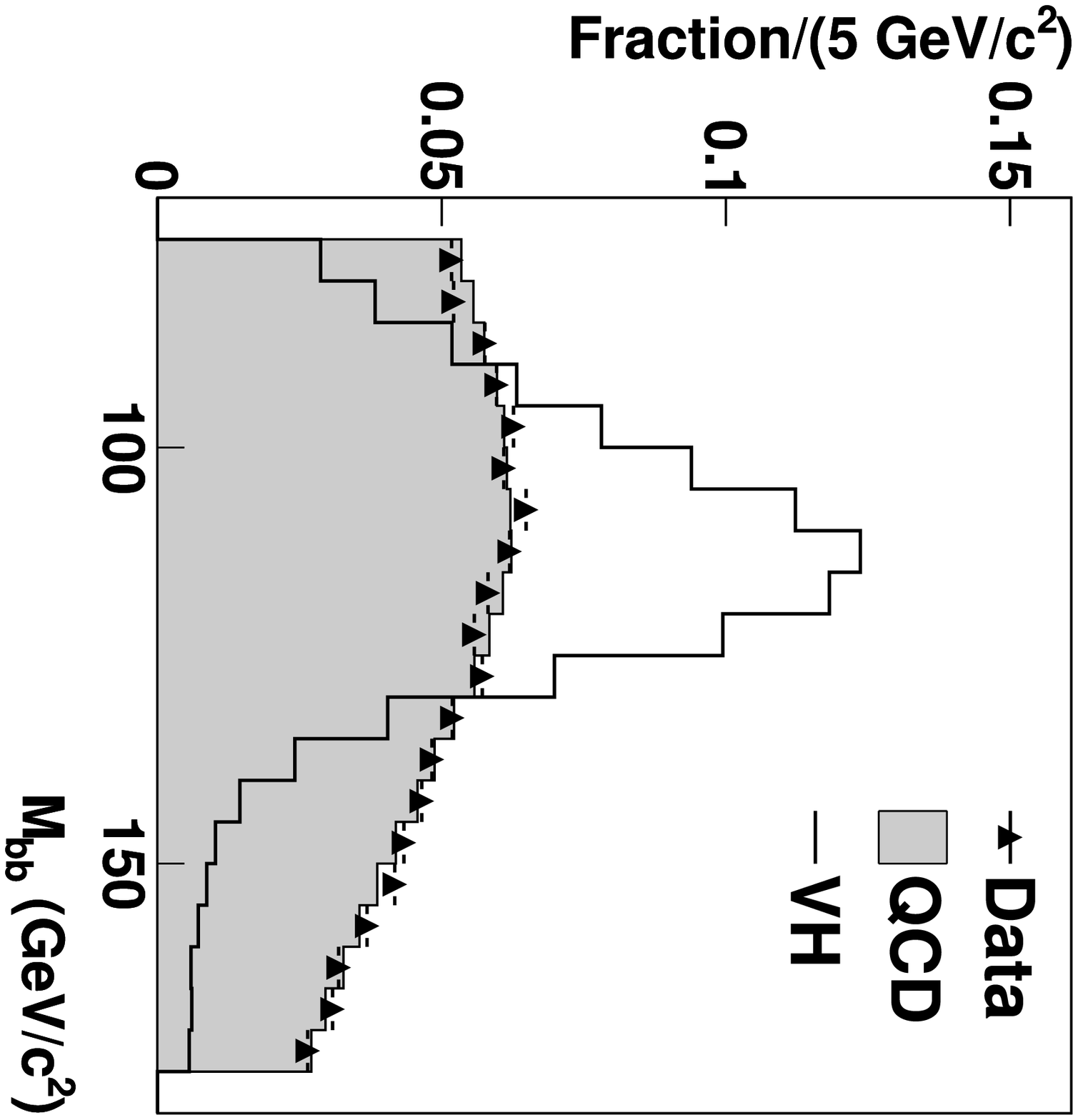}
}
%
\subfloat[] {
\includegraphics[width=6cm, angle=90]{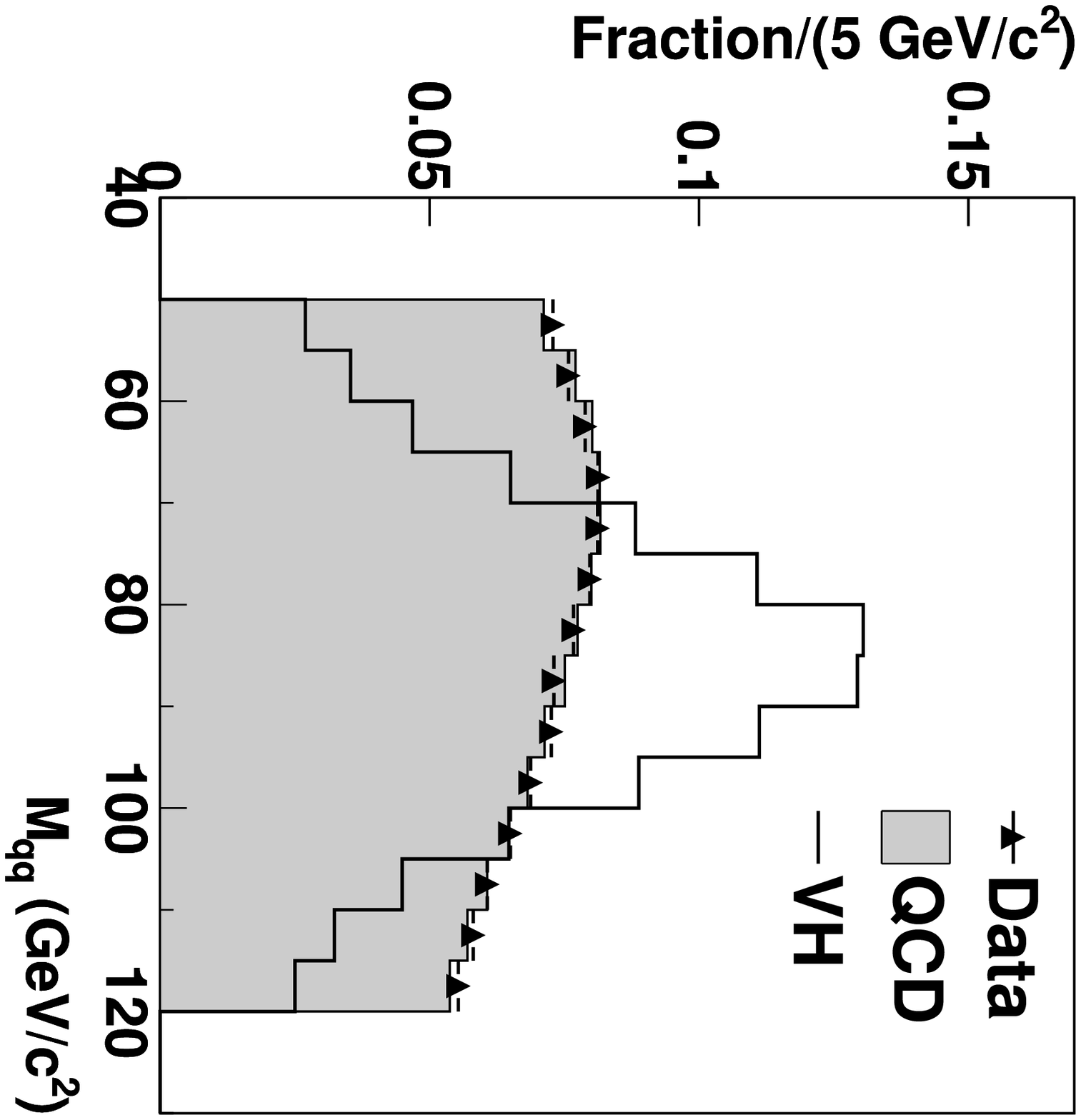}
}
%
\subfloat[] {
\includegraphics[height=6cm, angle=0]{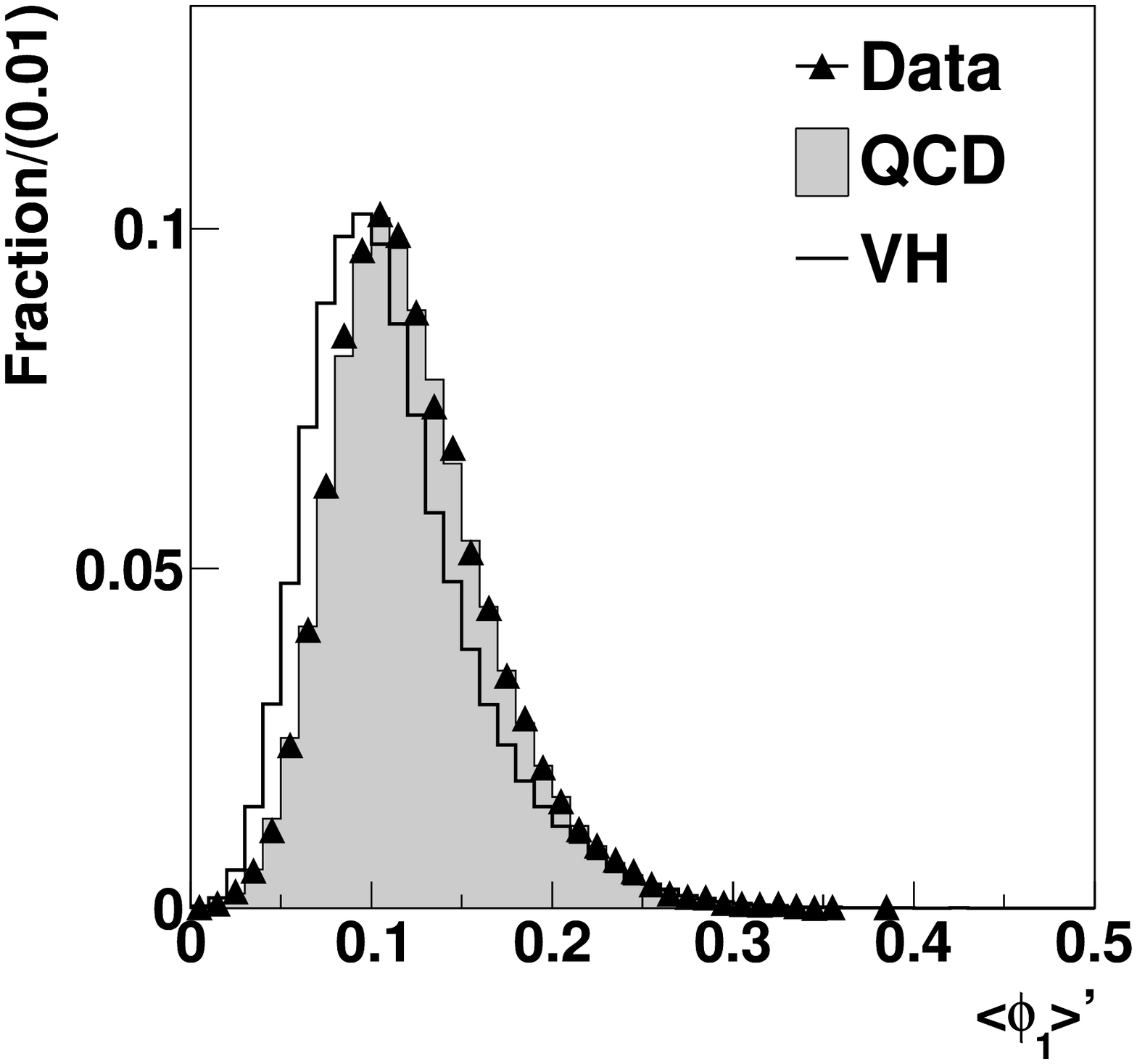}
}

\subfloat[] {
\includegraphics[height=6cm, angle=0]{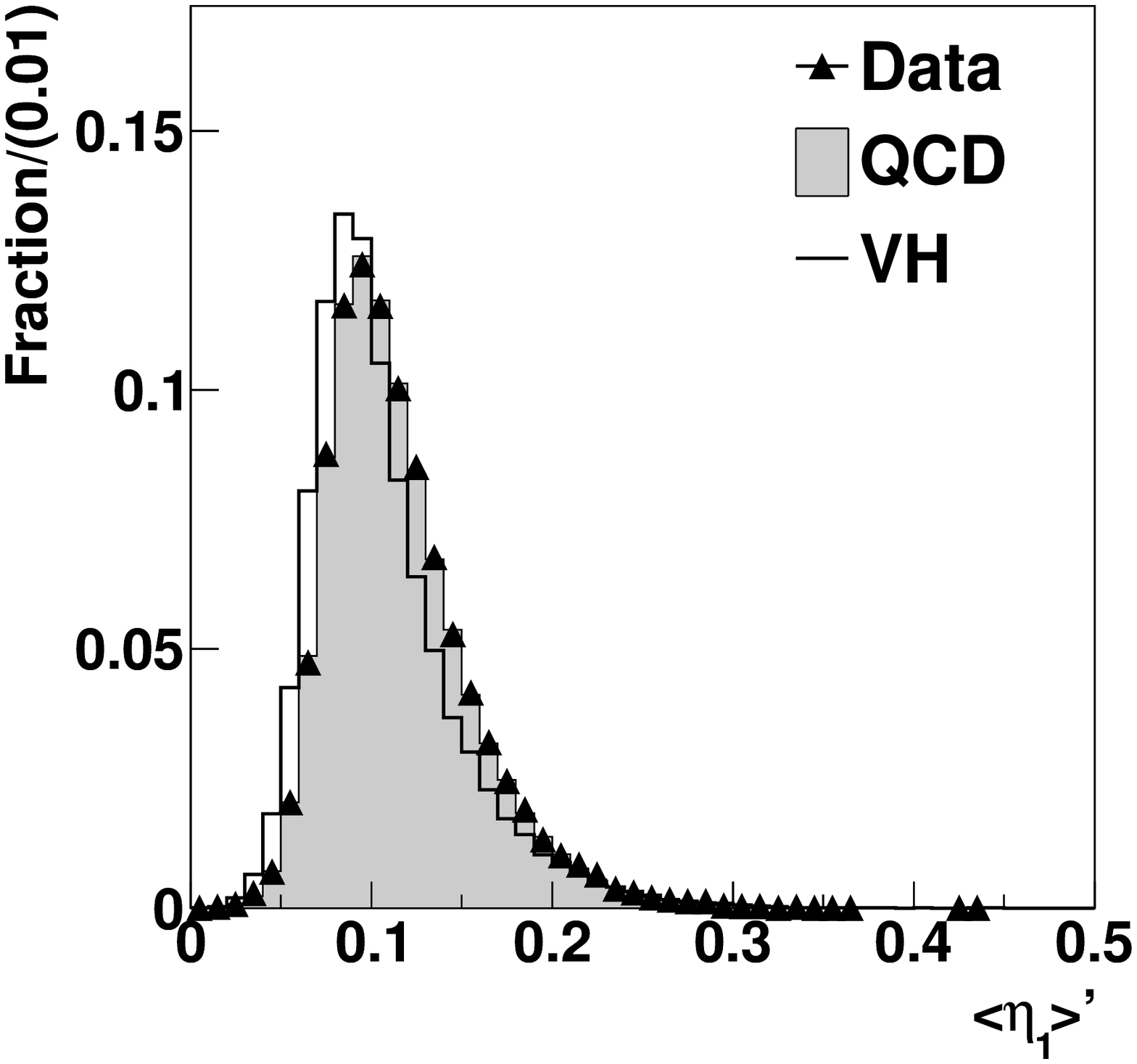}
}
%
\subfloat[] {
\includegraphics[height=6cm, angle=0]{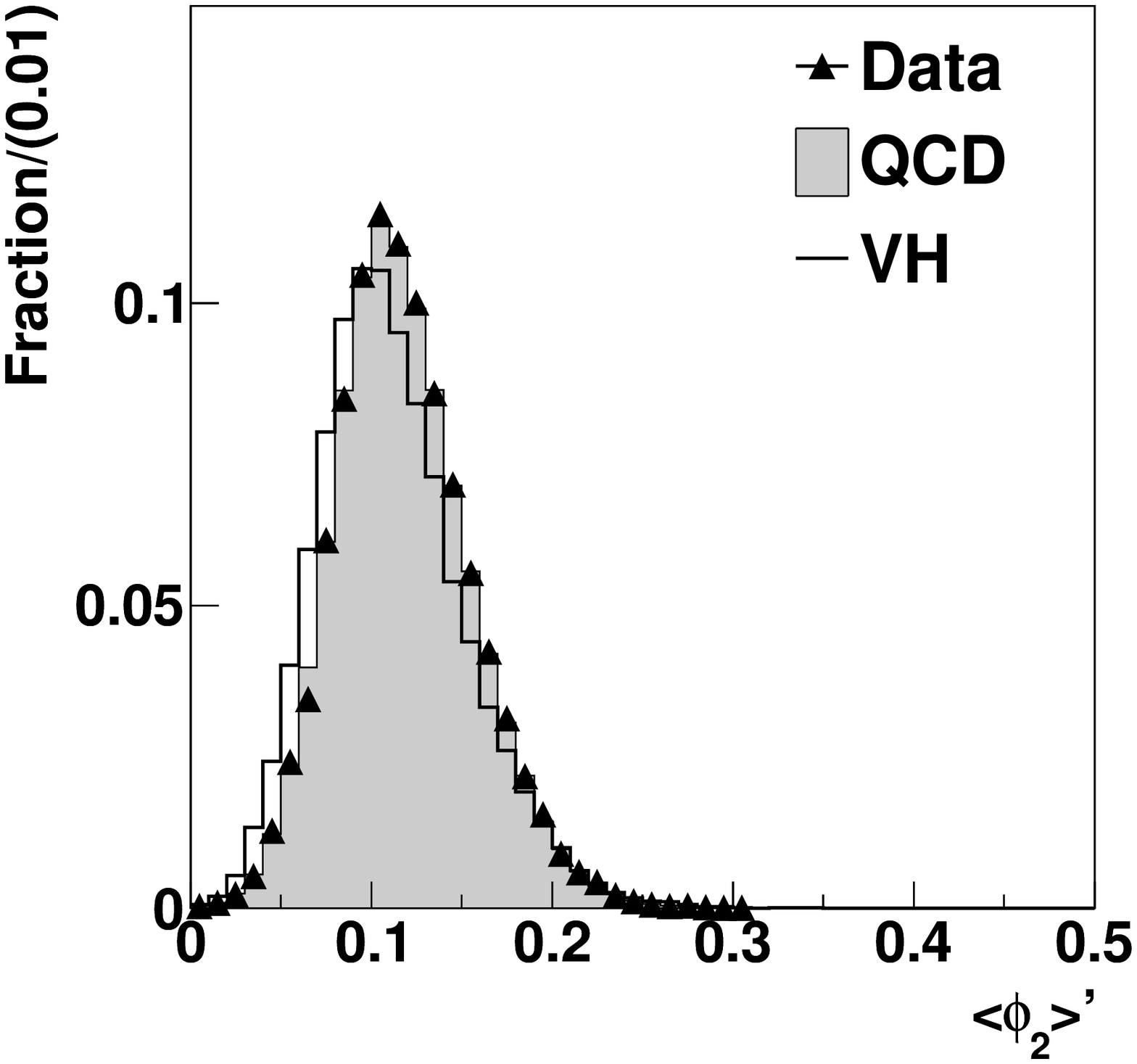}
}
%
\subfloat[] {
\includegraphics[height=6cm, angle=0]{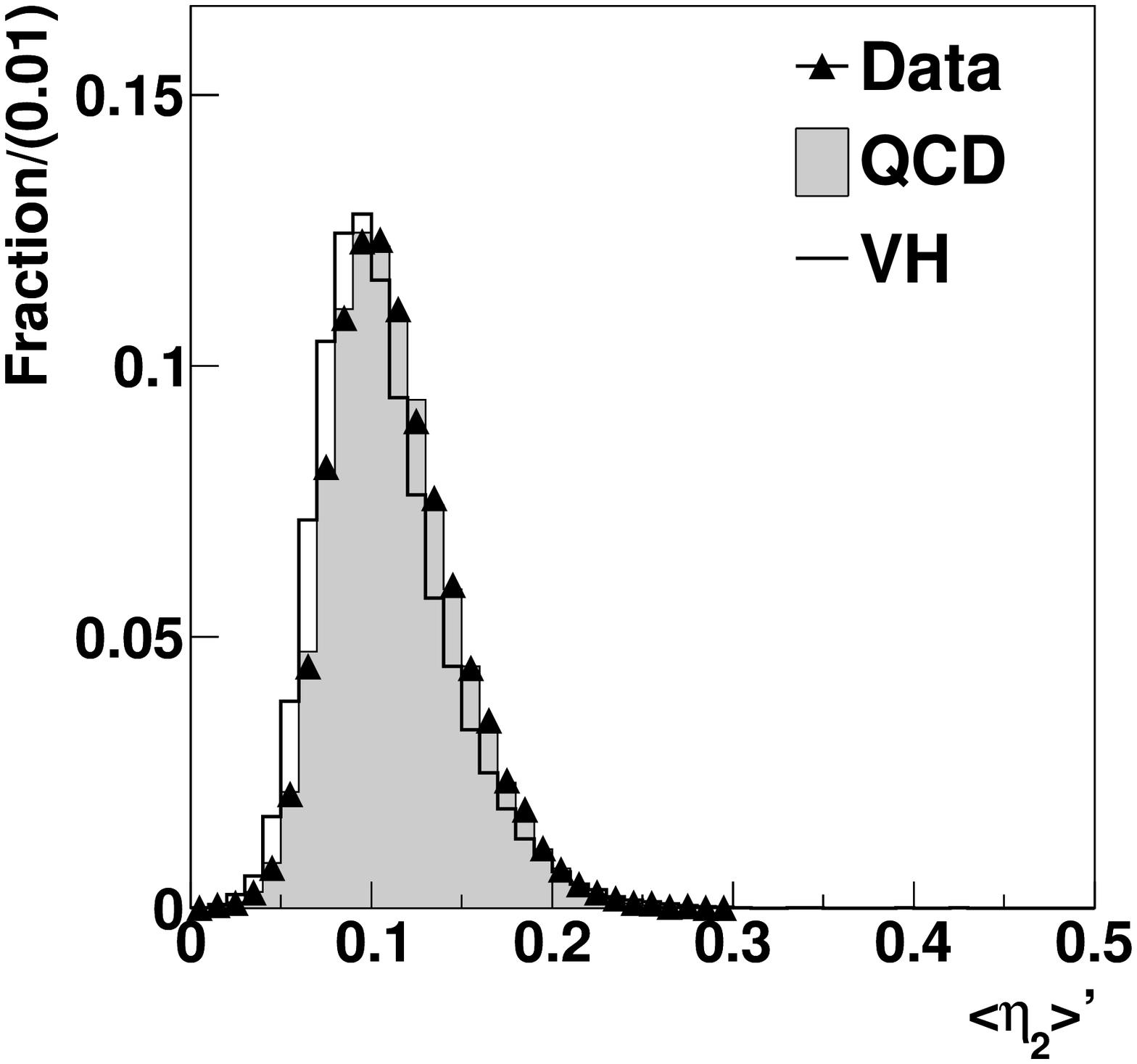}
}

\subfloat[] {
\includegraphics[width=6cm, angle=90]{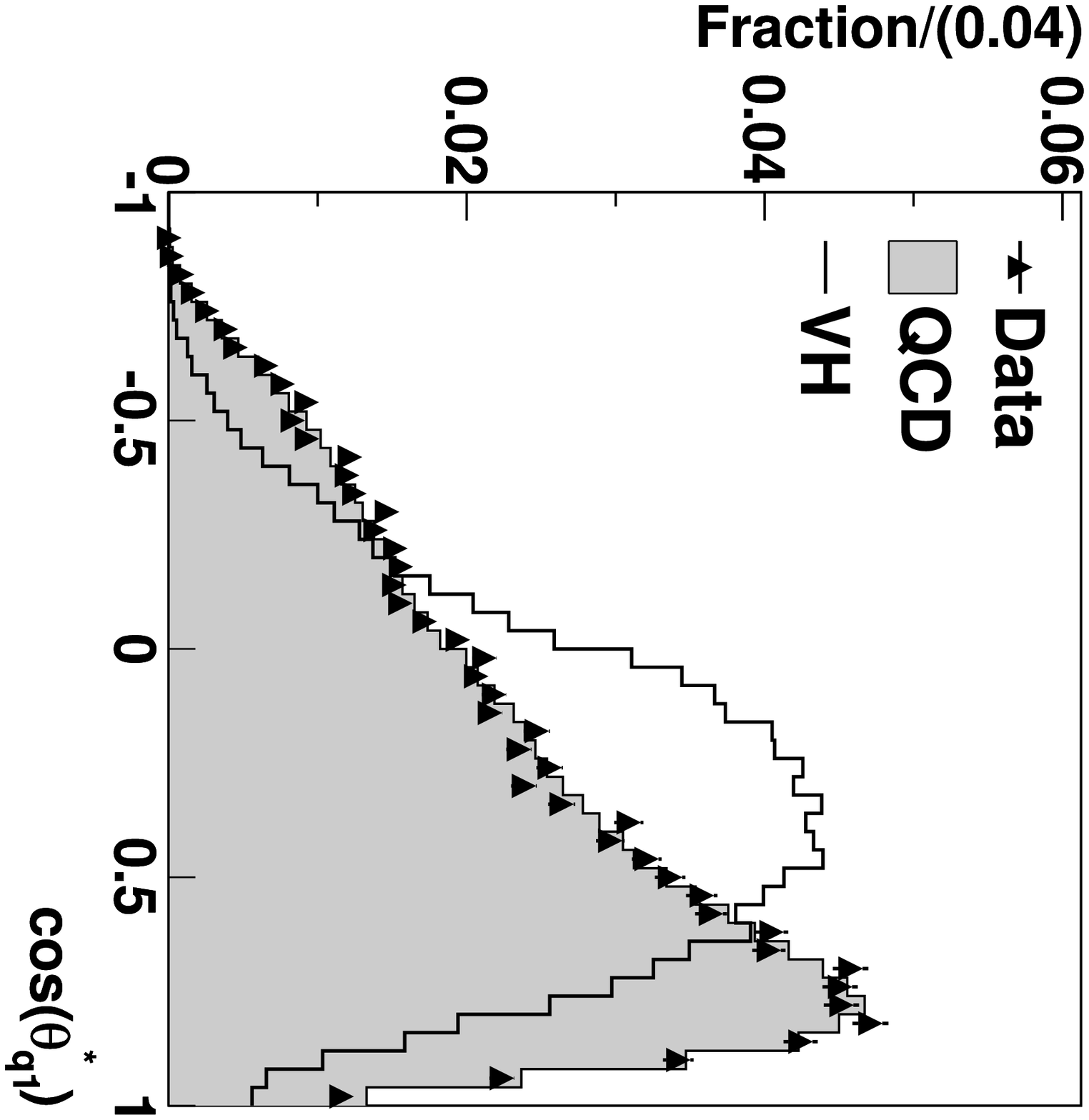}
}
%
\subfloat[] {
\includegraphics[width=6cm, angle=90]{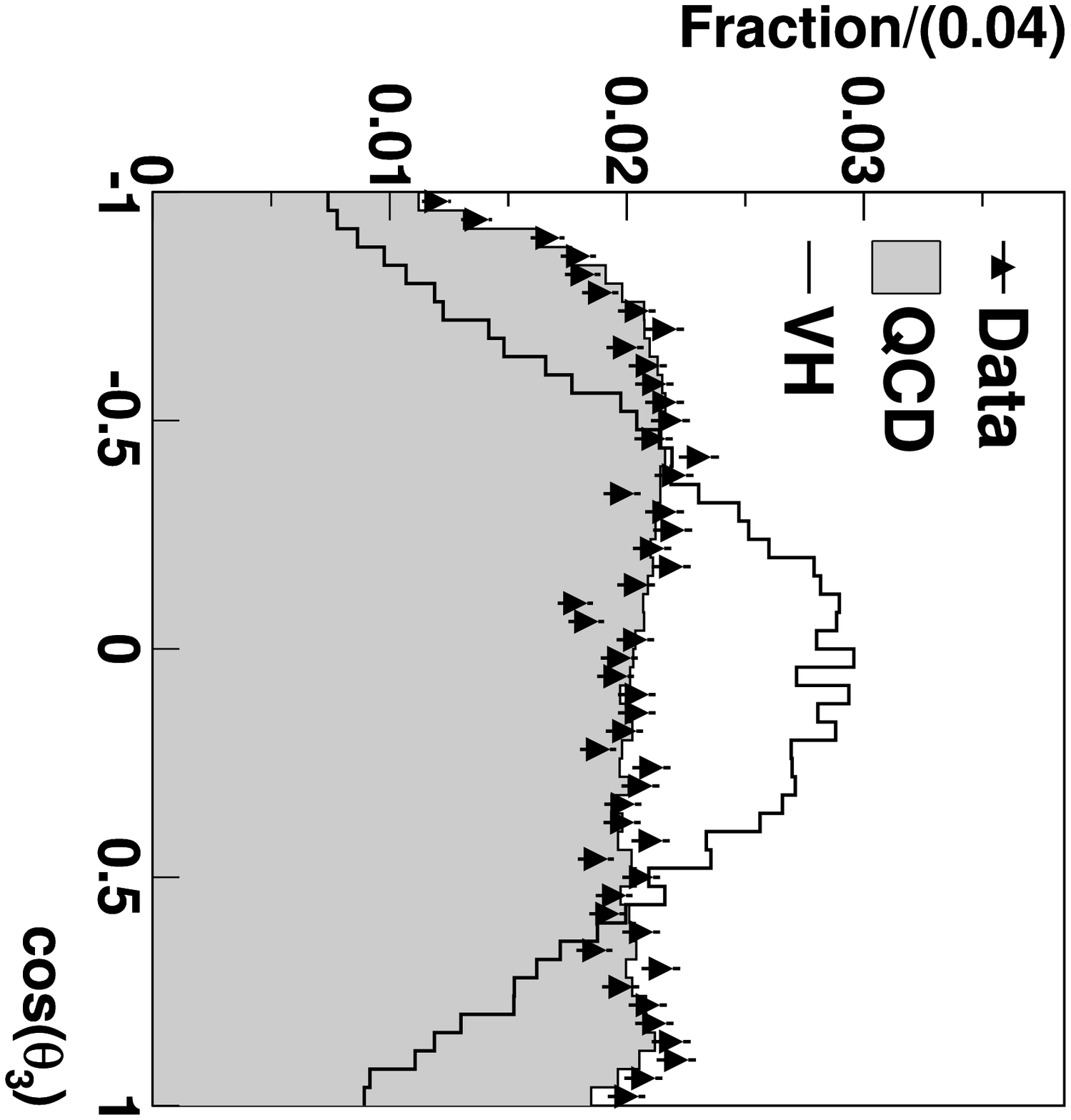}
}
%
\subfloat[] {
\includegraphics[width=6cm, angle=90]{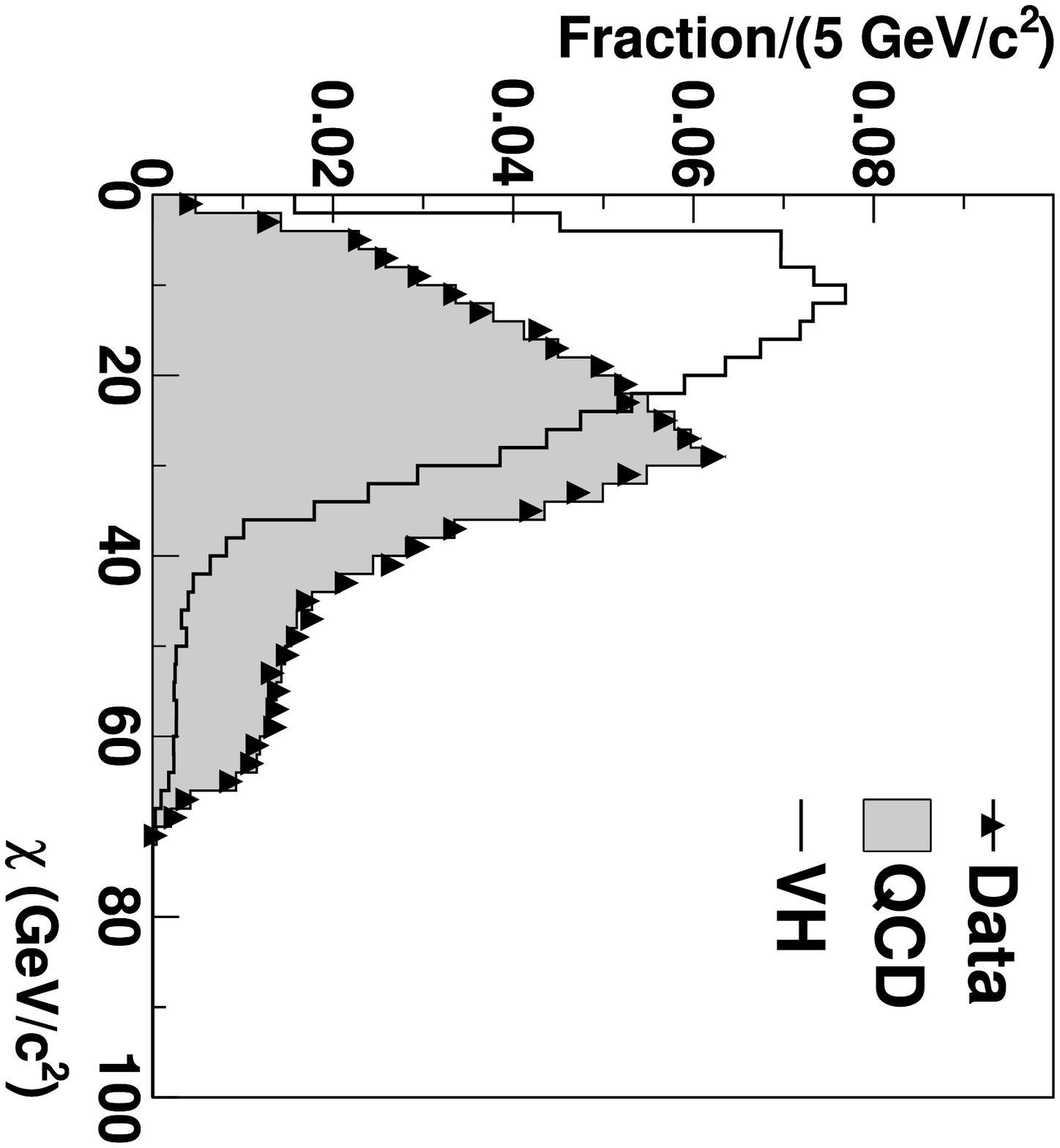}
}
\caption{Distribution of variables used to train the {\sl VH} NN. The signal consists of {\sl VH} ($m_{H}=120\,\gevcc$) SS events and the background consists of TRF predicted QCD SS events which have passed the {\sl VH} candidate selection.  All plots are normalised to unit area to compare shapes. After correction functions have been applied to \mqq\, and the jet-moments, the TRF correctly predicts the shape of the double-tagged SS data for all variables}
\label{FIG:VHNNInputVariables}
\end{figure*}

\begin{figure*}[p!]
%
\subfloat[] {
\includegraphics[width=6cm, angle=90]{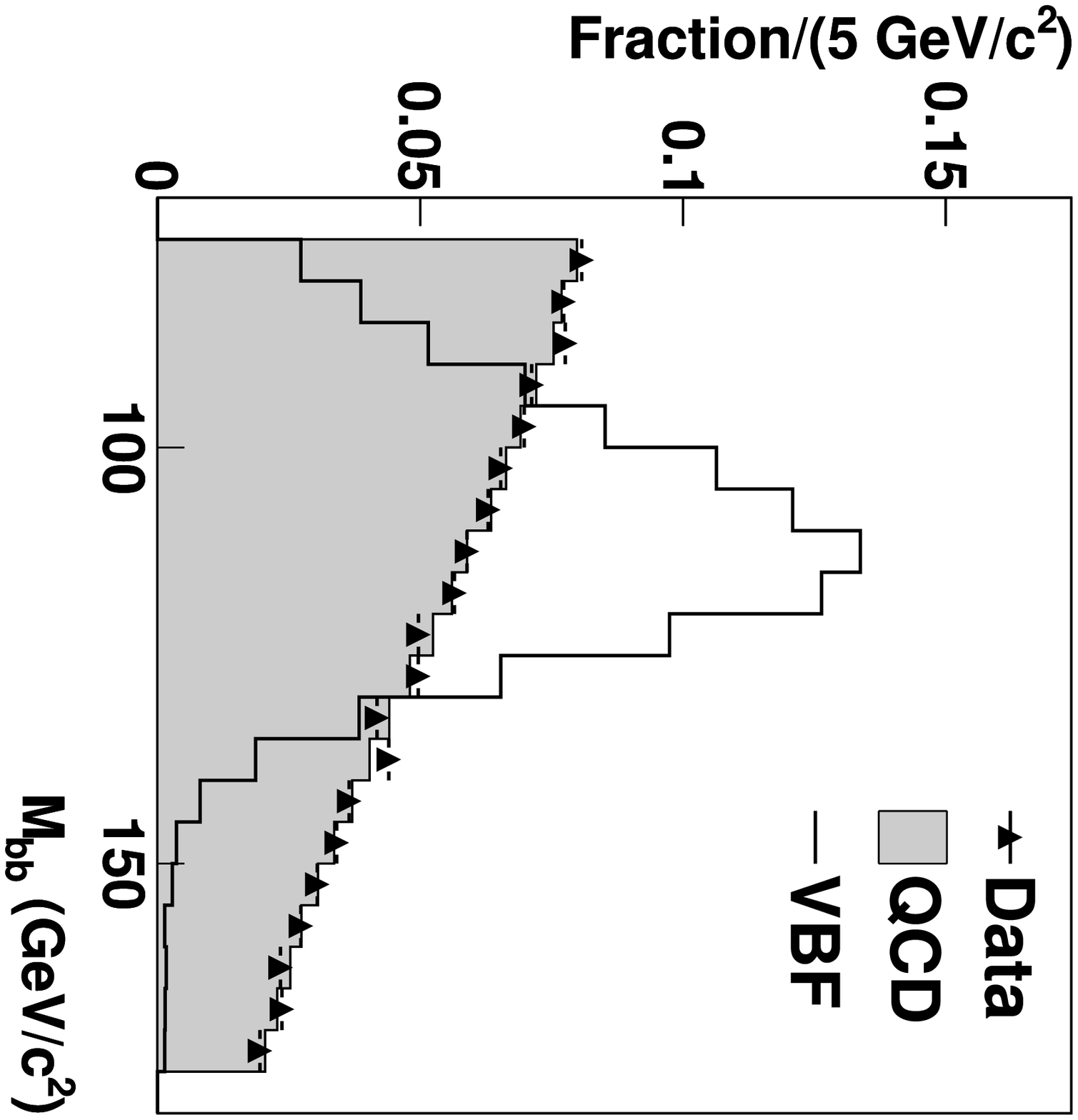}
}
%
\subfloat[] {
\includegraphics[width=6cm, angle=90]{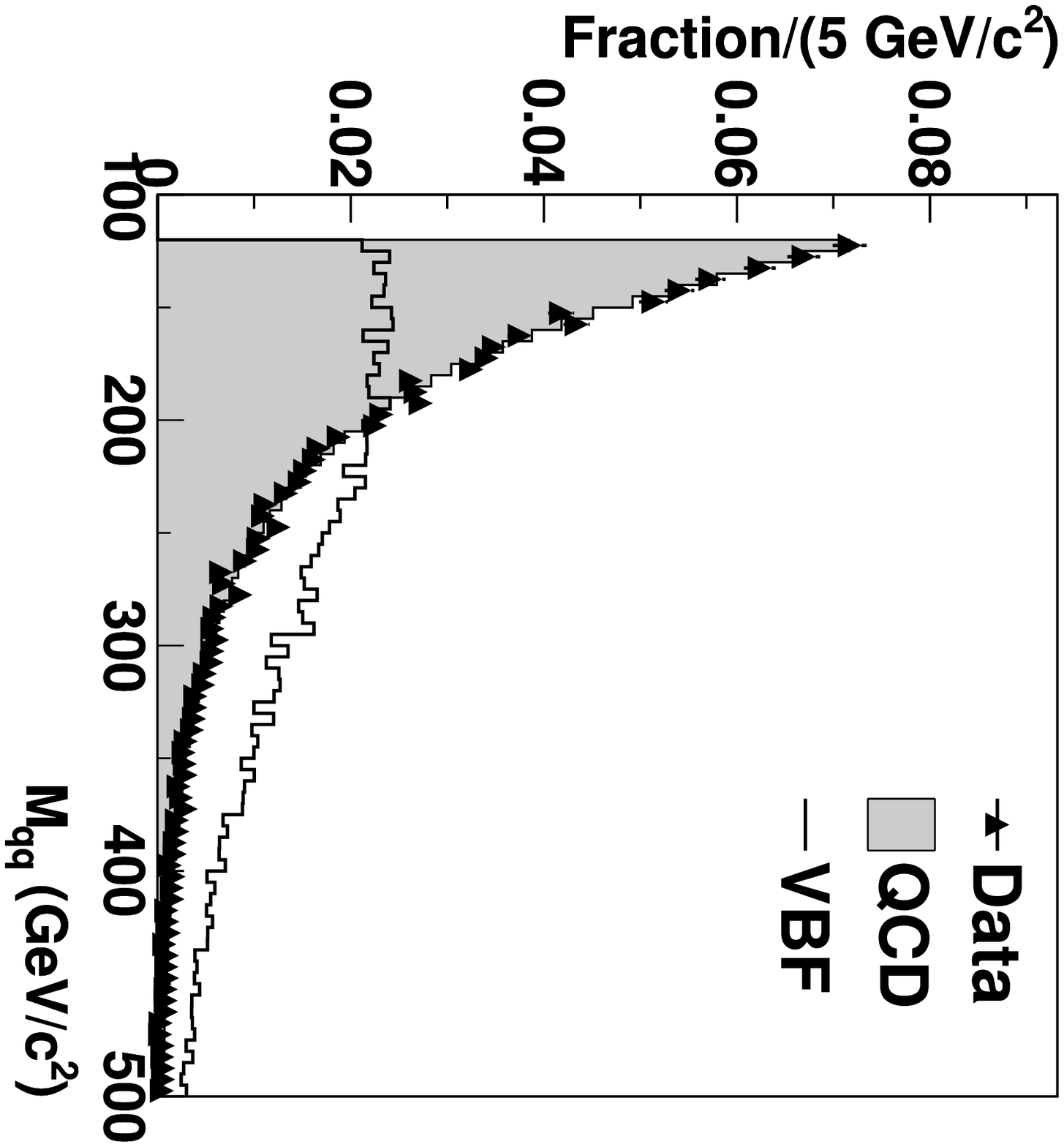}
}
%
\subfloat[] {
\includegraphics[height=6cm, angle=0]{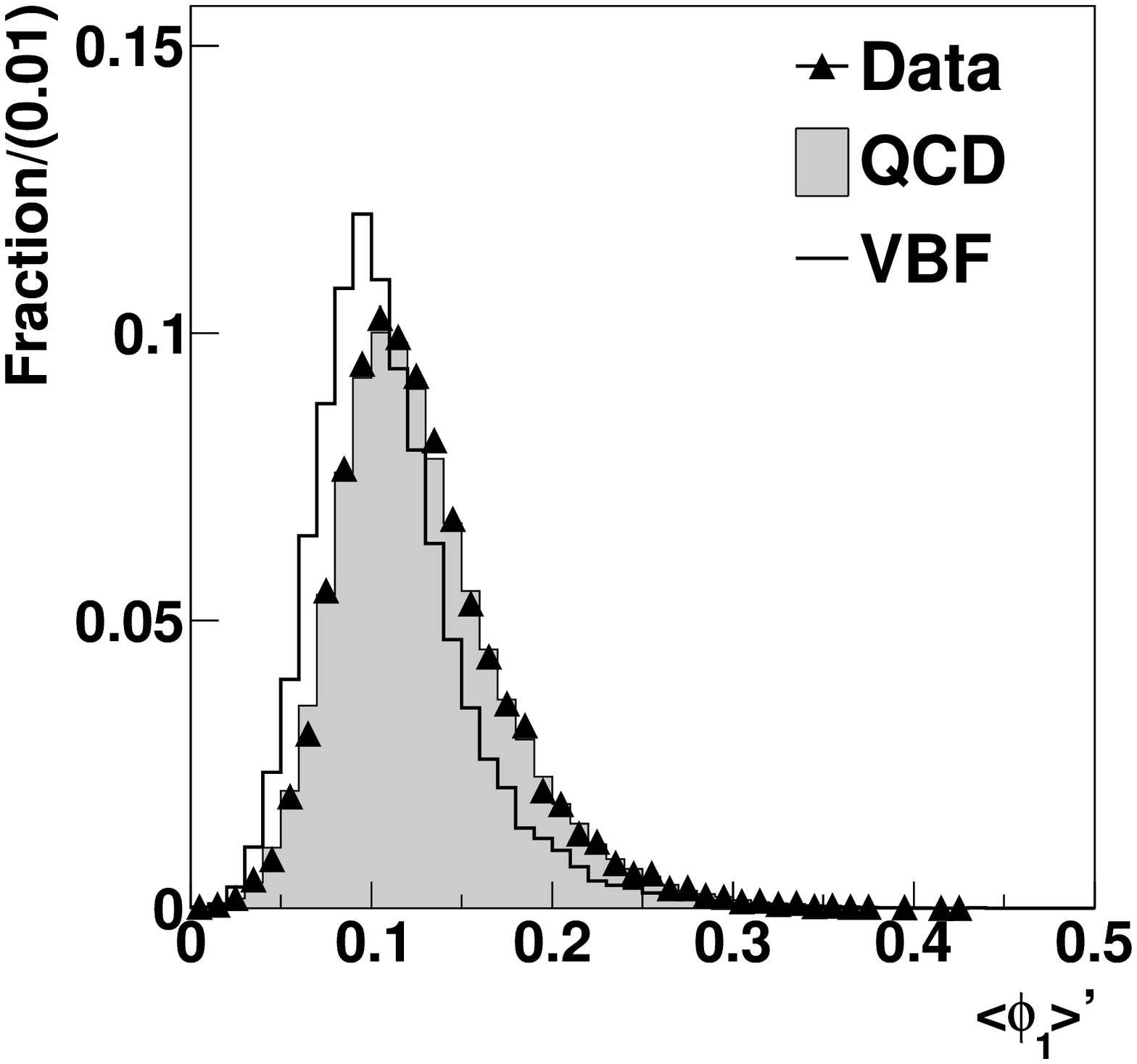}
}

\subfloat[] {
\includegraphics[height=6cm, angle=0]{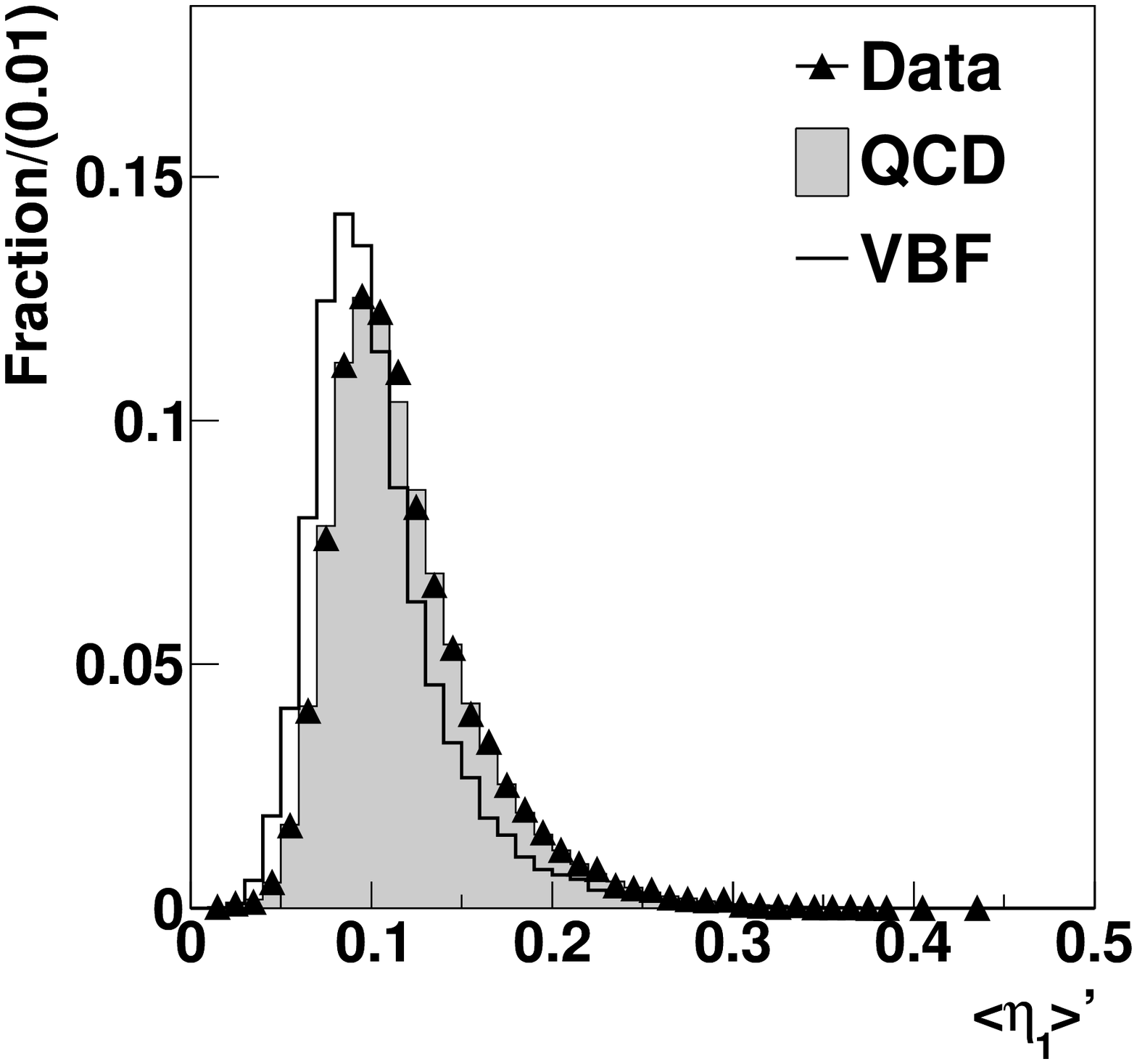}
}
%
\subfloat[] {
\includegraphics[height=6cm, angle=0]{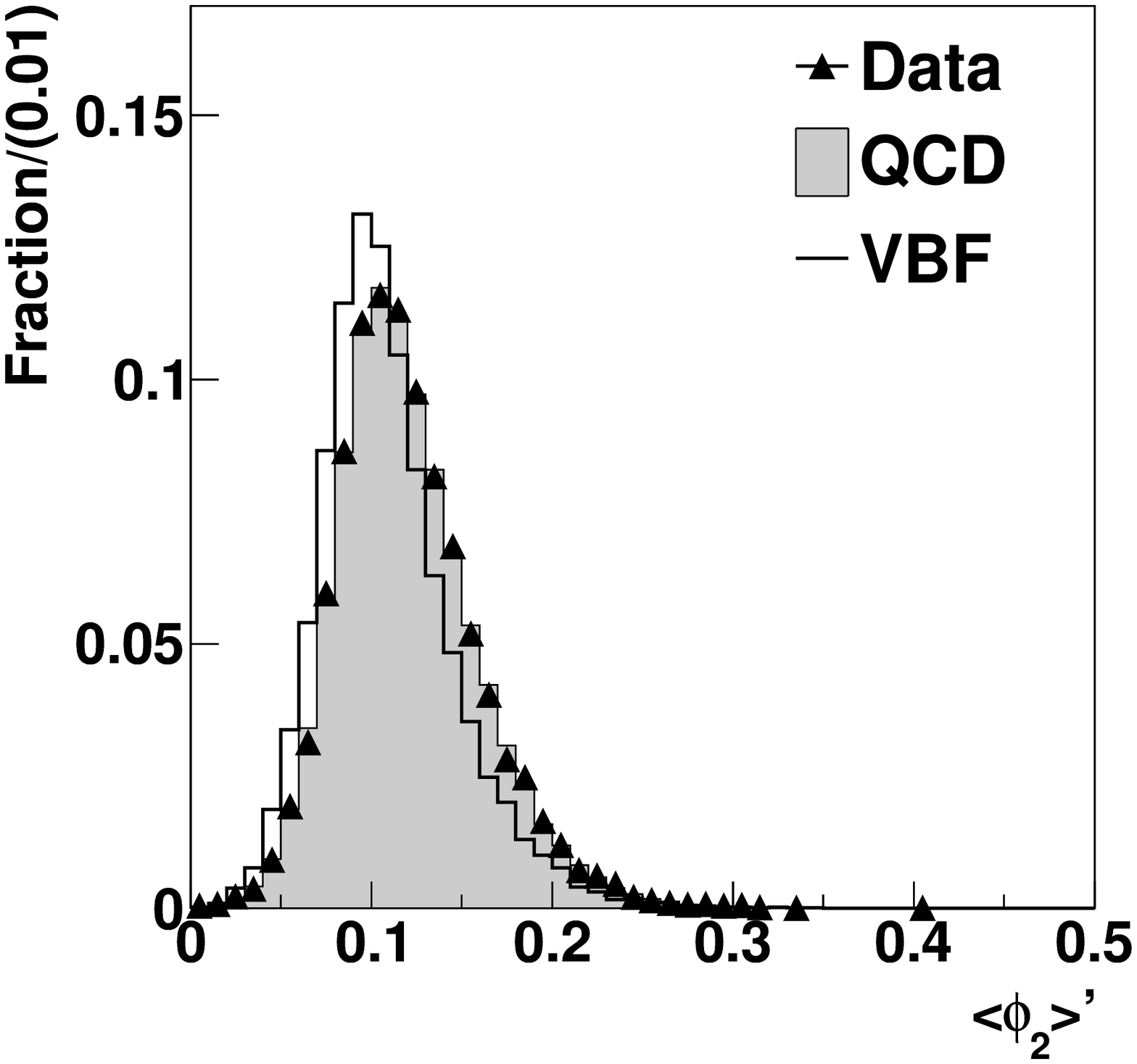}
}
%
\subfloat[] {
\includegraphics[height=6cm, angle=0]{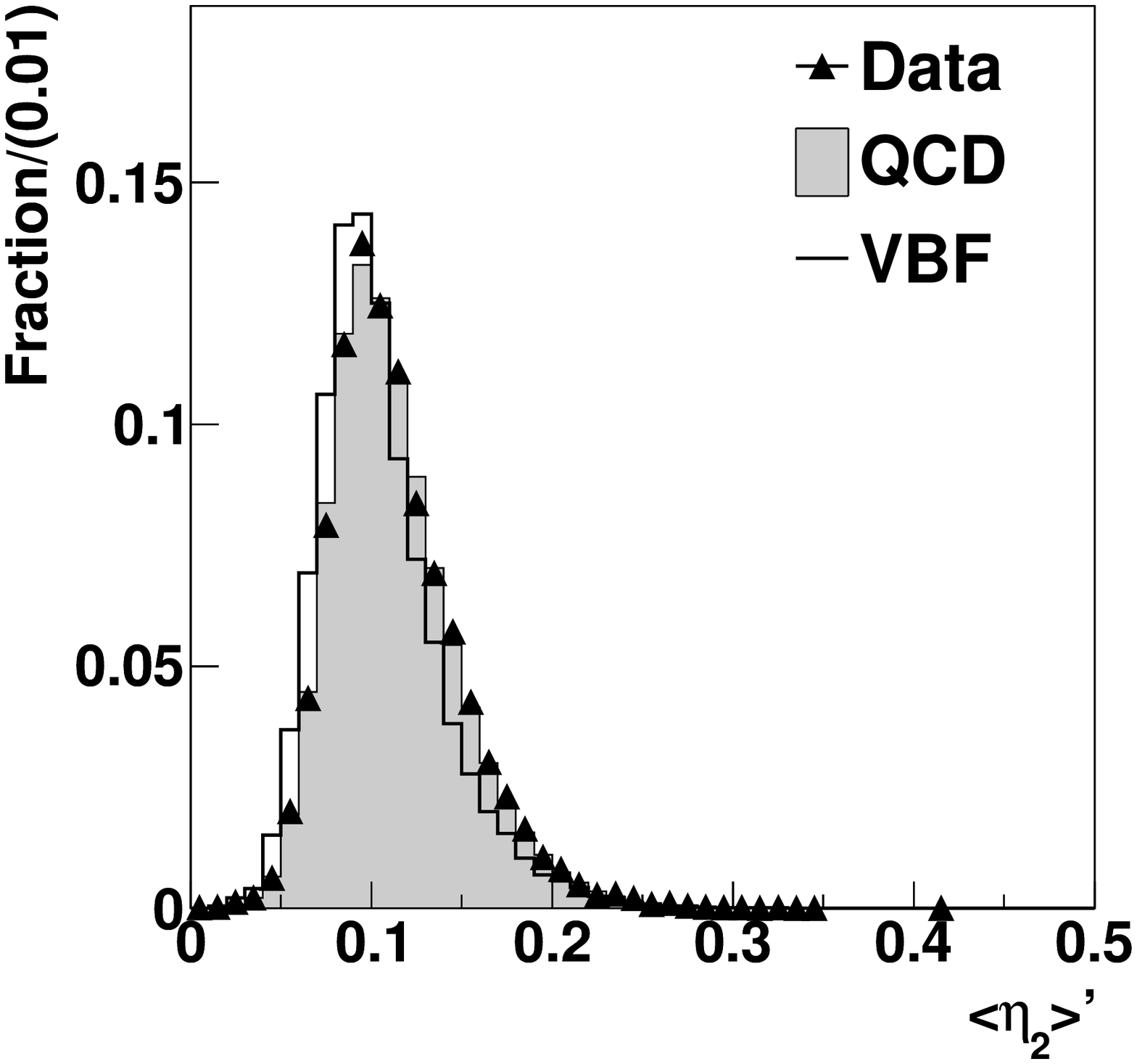}
}
\caption{Distribution of variables used to train the VBF NN. The signal consists of VBF ($m_{H}=120\,\gevcc$) SS events and the background consists of TRF predicted QCD SS events and which have passed the VBF candidate selection. All plots are normalised to unit area to compare shapes. After correction functions have been applied to \mqq\, and the jet-moments, the TRF correctly predicts the shape of the double-tagged SS data for all variables.}
\label{FIG:VBFNNInputVariables}
\end{figure*}

Before training the NN,  the TRF QCD modeling was verified by comparing the shapes of the NN training variables
constructed from single-tagged events, after applying the TRF,  and double tagged events from the TAG region. The TRF was able to reproduce the shapes of all the NN training variables except \mqq\, and the jet moments.  The TRF was corrected using a correction function for each mismodeled variable. The correction function was constructed from the fitted ratio of the observed double tagged shape in the {\sc Tag} region to the TRF prediction in the {\sc Tag} region.  The largest correction value was 2\%.  Figures~\ref{FIG:VHNNInputVariables} and~\ref{FIG:VBFNNInputVariables}  show distributions of the NN training variables of {\sl VH} and VBF signal,  corrected TRF QCD  and double tagged data for the SS $b$-tagging category. The corrected TRF QCD follows the shape of the data for all variables. The TRF predictions for SJ were validated in the same way.


We search for a Higgs boson of mass $100 \leq m_{H} \leq 150$\,\gevcc\, at 5\,\gevcc\, intervals.  As \mbb\, is one of the NN training variables, which varies with different Higgs mass hypotheses, the Higgs search sensitivity can be improved by training the NN at different Higgs masses. There is a separate {\sl VH}(VBF) NN trained at $m_{H} = $100\,\gevcc, 120\,\gevcc, and 140\,\gevcc.  For Higgs mass hypotheses between 100\,\gevcc\, and 110\,\gevcc, the NN trained with $m_{H} =100\,\gevcc\,$ is used. Similarly, we use the   $m_{H} =120\,\gevcc$ trained NN to search for a Higgs boson between 115\,\gevcc\, and 130\,\gevcc\, and the   $m_{H} =140\,\gevcc$ trained NN to search for a Higgs boson between 135\,\gevcc\, and 150\,\gevcc.

  

\begin{figure*}[!]
\subfloat[] {
\includegraphics[width=7cm]{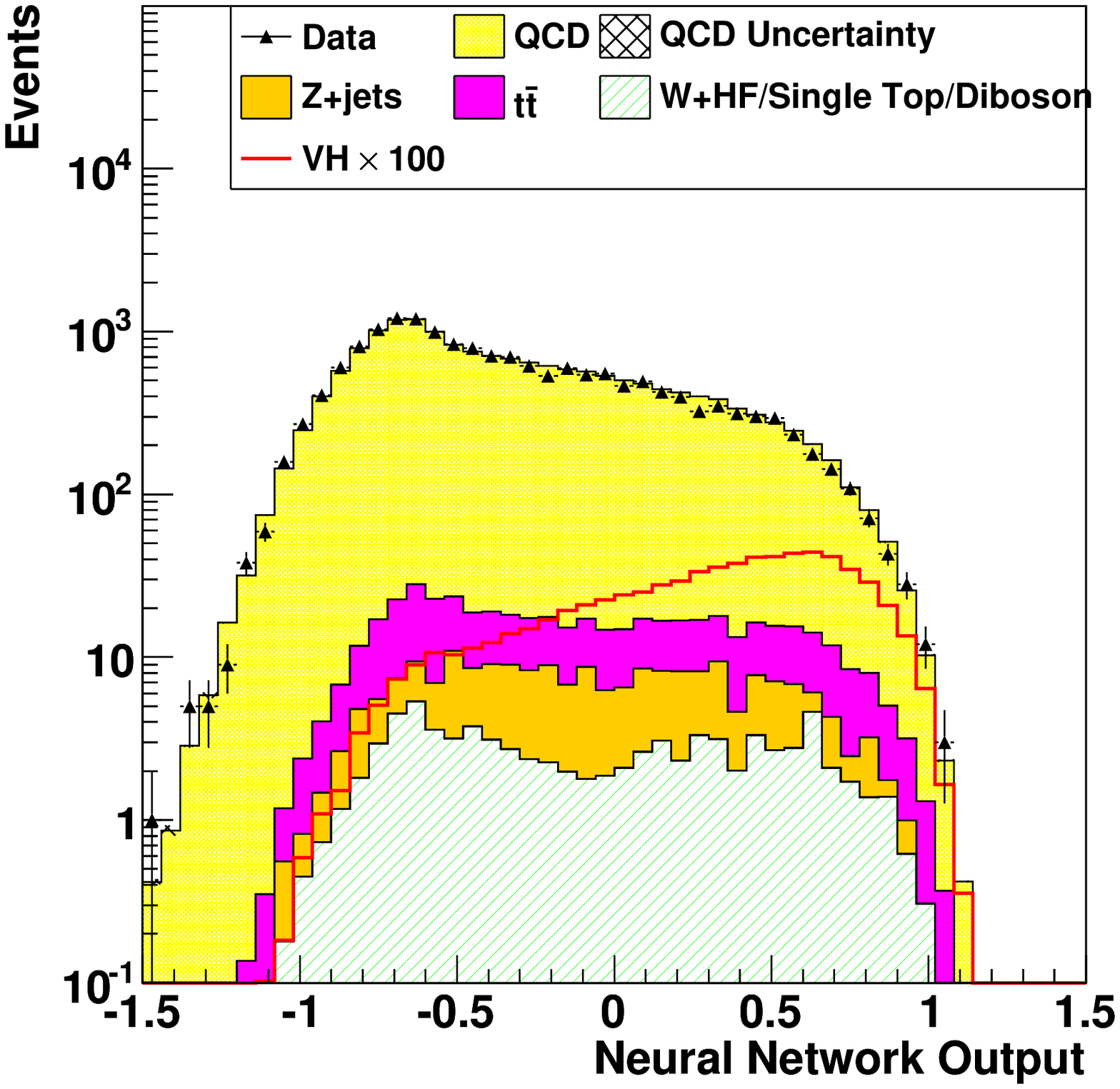}
}
\subfloat[ ] {
\includegraphics[width=7cm]{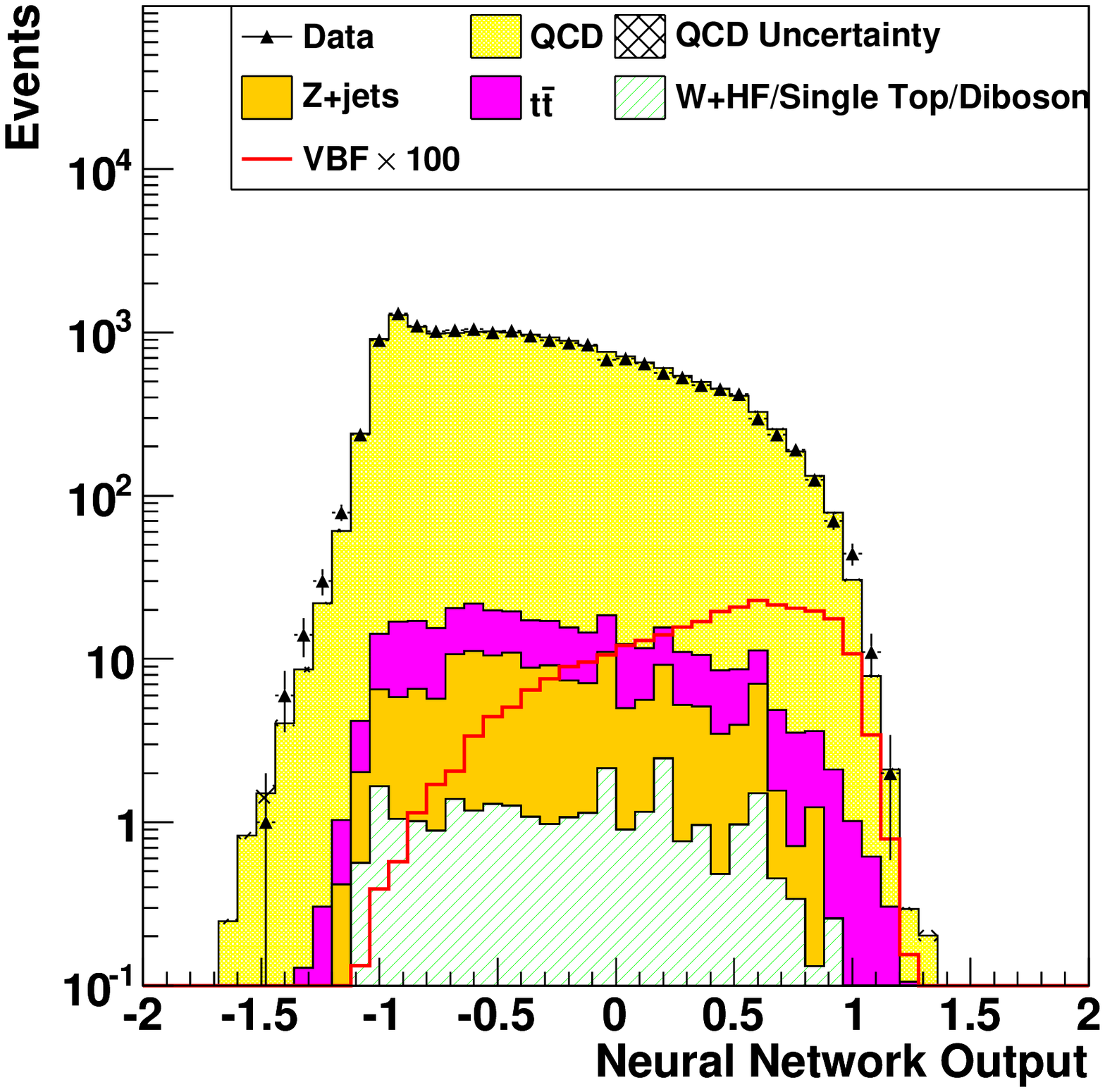}
}

\subfloat[] {                                 
\includegraphics[width=7cm]{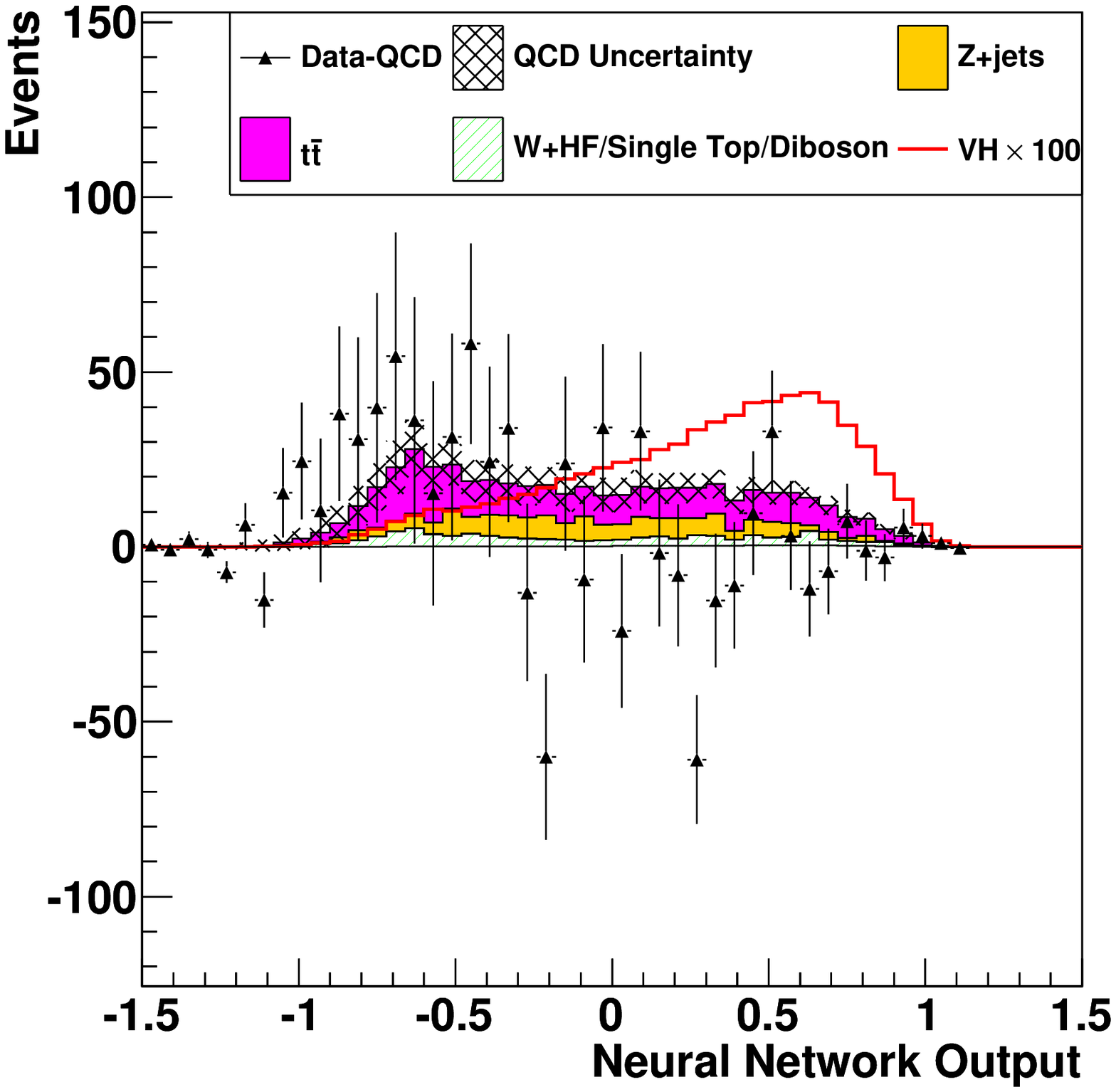}
}
\subfloat[] {
\includegraphics[width=7cm]{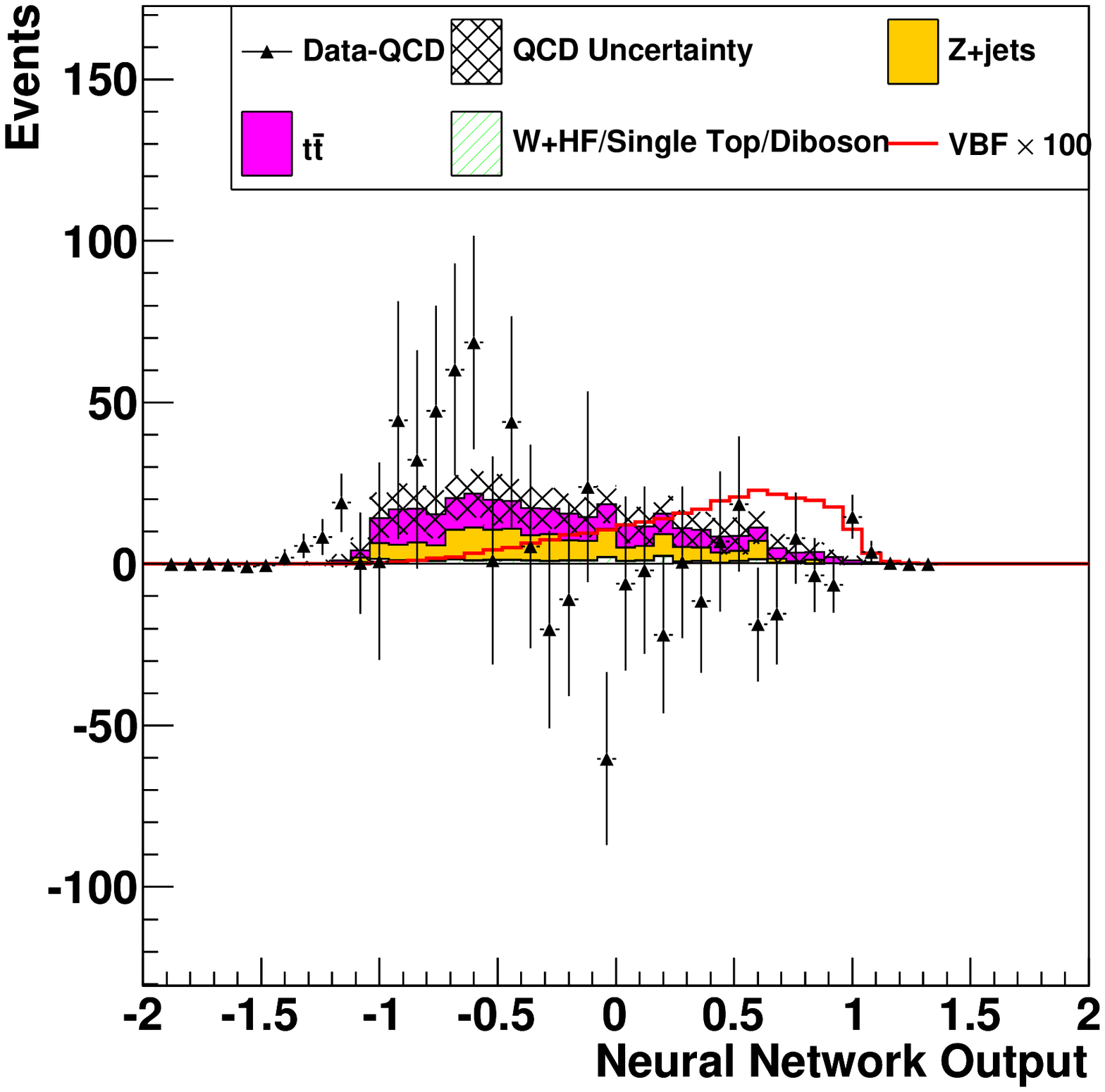}
}
\caption{\label{FIG:NNPlots}NN distribution for {\sl VH}-SS\,(a) and VBF-SS\,(b) for $m_{H} = 120\,\gevcc$.  
As the QCD background is large, data-QCD versions for {\sl VH}-SS and VBF-SS are shown in (c) and (d), respectively. The {\sl VH} and VBF distributions are scaled by a factor of 1000 and 100 for (a), (b)  and  (c), (d), respectively.} 
\end{figure*}

Figure~\ref{FIG:NNPlots} shows the NN distributions for {\sl VH} and VBF for a Higgs mass of 120\,\gevcc. The NN returns a more negative (positive)  score for background (signal) events.  As the QCD background is large, QCD subtracted NN distributions are also shown.


\section{Systematic Uncertainties}

We estimate the effect of systematic uncertainties by propagating uncertainties on the NN input variables to the NN output.  We consider both variations on the normalization and shape of the NN output.  The systematic uncertainties which affect the normalization of the Higgs signal and non-QCD backgrounds are: jet energy scale (JES)~\cite{Bhatti:2005ai}, parton distribution function (PDF) , $b$-tagging scale factor between MC and data, initial and final state radiation (ISR/FSR), trigger efficiency, integrated luminosity and cross sections~\cite{Collaboration2010Combined-CDF-an}. The Higgs signal cross section uncertainty is $\approx 5\%$~\cite{Collaboration2010Combined-CDF-an}
but is not included as its effect is negligible. 

The uncertainties which affect the Higgs signal NN output shape  are the JES, ISR/FSR, and jet moments. Sec.~\ref{SEC:JetMoment} defined the jet moment uncertainty as half of the offset required to correct the MC. The Higgs signal jet moment NN shape uncertainty is defined as the difference of the Higgs signal NN shape using the nominal jet moment correction and the Higgs signal NN shape using half of the jet moment correction. 

For the TRF QCD prediction, we consider two shape uncertainties: an interpolation uncertainty and correction function uncertainty for \mqq, $\langle \phi \rangle$, and $\langle \eta \rangle$ variables. The interpolation uncertainty accounts for possible difference in the TRF between the regions where it was measured ({\sc Tag}) and applied ({\sc Signal}) (Fig.~\ref{FIG:MbbMqqPlane}). An alternative TRF was measured using events in the {\sc Control} region, as indicated in Fig.~\ref{FIG:MbbMqqPlane}, which is still background-dominated. The interpolation uncertainty is 
defined as the  difference of the QCD NN shapes using the nominal {\sc Tag} TRF and {\sc Control} TRF. The correction function uncertainty for \mqq\, is evaluated by deriving an alternative correction function for \mqq\, 
using events from the {\sc control} region.  This alternative \mqq\, correction function is applied to the TRF, instead of the nominal \mqq\, correction function, and propagated through the NN.  The difference in the QCD NN shape between using the nominal correction function and the {\sc Control} region derived function defines the correction function shape uncertainty.  The systematic uncertainty for  $\langle \phi \rangle$, and $\langle \eta \rangle$ correction functions is evaluated in the same way.  The QCD NN output varied at most by $\sim 2-3\%$ for each QCD shape systematic. The uncertainties are summarized in Table~\ref{TABLE:Systematics}.

\begingroup
\squeezetable
\begin{table}
\caption{Summary of systematic uncertainties. The largest  change is quoted for the sources which have a shape uncertainty.}
\begin{ruledtabular}
\begin{tabular}{l l}
Source                &      Uncertainty   \\ \hline
\multicolumn{2}{l}{Higgs and non-QCD uncertainties} \\
Integrated Luminosity & $\pm 6\%$ \\ 
Trigger efficiency          & $\pm 4\%$ \\
PDF                                &  $\pm 2\%$ \\
JES                                 &  $\pm 7\%$  and shape\\
$b$-tagging scale factor & $\pm 7.6\%$ for SS \\
                                        &  $\pm 9.7\%$ for SJ \\
ISR/FSR                        &  $\pm 2\%$ and shape for {\sl VH} \\
                                       &   $\pm 3\%$ and shape for VBF \\
$\langle \phi	\rangle^{\prime}$ and $\langle \eta	\rangle^{\prime}$ & shape for {\sl VH} and VBF  $(\leq 10\%)$    \\
%
%
\ttbar\, and single-top cross section & $\pm 10\%$ \\
Diboson cross section  & $\pm 6\%$ \\
$W$+HF \& $Z$+Jets cross section & $\pm 50\%$ \\  \hline
\multicolumn{2}{l}{QCD uncertainties} \\
interpolation                                                       & shape  $(\leq 3\%)$  \\
\mqq\, correction function                                  & shape $(\leq 1\%)$  \\
$\langle \phi \rangle$ correction function    & shape   $(\leq 2\%)$ \\
$\langle \eta \rangle$ correction function    & shape   $(\leq 2\%)$  \\                                         
\end{tabular}
\end{ruledtabular}
\label{TABLE:Systematics}
\end{table}
\endgroup



\section{Results}


The NN output distribution of data are compared to the background and we find no excess of events over the expected background. We calculate upper limits on the excluded Higgs boson cross section at the  95\% CL for Higgs boson mass hypotheses  $100 \leq m_{H} \leq 150$\,\gevcc\, at 5\,\gevcc\, intervals. 
%
%
The limits are calculated using a Bayesian likelihood method with a flat prior for the signal cross section.
We integrate over Gaussian priors for the systematic uncertainties and incorporate correlated rate and shape uncertainties as well as uncorrelated bin-by-bin statistical uncertainties~\cite{PhysRevLett.104.061802}. The QCD normalization is a free parameter that is fit to the data.

The median of the 95\% CL obtained from 10000 simulated experiments is taken at the expected 95\% CL. The $\pm 1 \sigma$ (where $\sigma$ denotes the standard deviation) and $\pm 2 \sigma$ expected limits are derived from the 16th, 84th, 2nd and 98th percentiles of the distribution, respectively. 

\begin{figure}[!]
\begin{center}
\includegraphics[width=7cm]{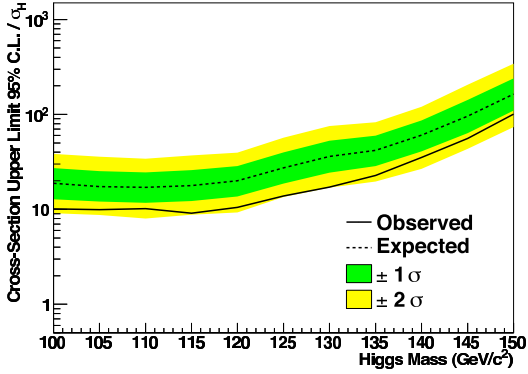}
\caption{Expected (dashed) and observed (solid) 95\% CL normalized to the SM cross section for the combined {\sl VH} and VBF channel. The dark (light) band represents the $1\sigma (2\sigma)$ expected limit range.}
\label{FIG:CombinationLimits}
\end{center}
\end{figure}

For $m_{H}$ = 120\,\gevcc, the observed (expected) limit, normalized to the SM cross section, for the individual analysis channels are 11.9(25.6) for {\sl VH}-SS, 43.4(51.8) for {\sl VH}-SJ, 47.0(49.4) for VBF-SS, 93.7(132.3) for VBF-SJ, and  10.5(20.0) for the combination of these four channels. The combined channel limits for Higgs boson masses in the range between 100 - 150\,\gevcc\, are shown in Fig.~\ref{FIG:CombinationLimits} and summarized in Table~\ref{TAB:MCLimits}.

%
%




\begingroup
\squeezetable
\begin{table}
\caption{Expected and observed 95\% CL upper limits for the combined {\sl VH} and VBF channels. The limits are normalized to the expected Higgs cross section.}
\begin{ruledtabular}
\begin{tabular}{c r r r r r r}
Higgs mass    & $-2\sigma$ & $-1\sigma$ & Median   & $+1\sigma$ & $+2\sigma$ & Observed \\
(GeV/c$^{2}$) &            &            &  &  &  &          \\ 
\hline
100 & 9.1    & 12.8   & 18.8   & 27.2    & 38.5   & 10.1   \\
105 & 8.7    & 12.1   & 17.4   & 25.2    & 35.8   & 9.9     \\
110 & 8.0    & 11.7   & 17.1   & 24.5    & 34.2   & 10.2   \\
115 & 8.8    & 12.2   & 17.8   & 25.9    & 36.9   & 9.1     \\
120 & 9.3    & 13.7   & 20.0   & 28.5    & 39.5   & 10.5   \\
125 & 13.5  & 18.7   & 27.3   & 39.8    & 57.0   & 13.8   \\
130 & 17.0  & 24.4   & 36.1   & 52.8    & 75.4   & 17.2   \\
135 & 19.6  & 28.6   & 41.9   & 59.7    & 82.7   & 22.7   \\
140 & 26.7  & 40.7   & 60.4   & 86.6    & 120.2 & 35.2   \\
145 & 43.4  & 63.5   & 95.7   & 142.1  & 205.3 & 55.8   \\
150 & 73.8  & 109.9 & 164.1 & 240.3 & 341.9 & 101.0 \\
\end{tabular}
\end{ruledtabular}
\label{TAB:MCLimits}
\end{table}
\endgroup

\begin{figure}
\subfloat[VH-SS] {
\includegraphics[width=5cm]{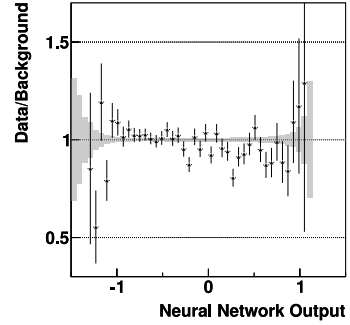}
}

\subfloat[VH-SJ] {
\includegraphics[width=5cm]{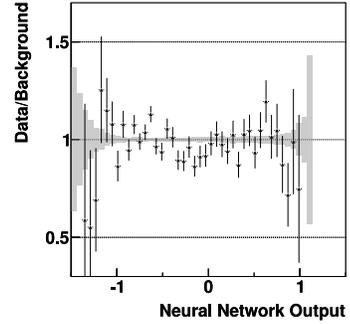}
}

\subfloat[VBF-SS] {
\includegraphics[width=5cm]{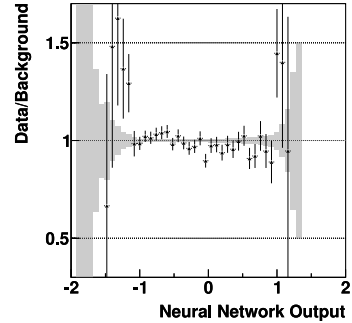}
}

\subfloat[VBF-SJ] {
\includegraphics[width=5cm]{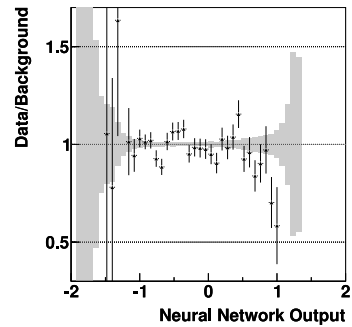}
}
\caption{Ratios of the data to background for {\sl VH}-SS (a), {\sl VH}-SJ (b), VBF-SS (c), and VBF-SJ (d) for the NN trained on 120\,\gevcc\, Higgs boson MC. The error bars of the data to background ratio are the statistical errors. The solid gray band is the ratio of the background systematic uncertainty  to the background.}
\label{FIG:DataBackgroundRatio}
\end{figure}

The observed limits for the individual search channels agree within $1\sigma$ of their expected limit except for the {\sl VH}-SS channel where we see a $2\sigma$ discrepancy.  The observed data in the {\sl VH}-SS channel have a deficit in the high signal region of the NN. Since the {\sl VH}-SS channel is the most sensitive, it has the strongest influence on the combined limit; thus the deviation of the observed limit from the expected limit in the VH-SS channel is similar to that of the combined limit. 
Figure~\ref{FIG:DataBackgroundRatio} shows the ratio of the data to the expected background for the four analysis channels for the NN trained on a 120\,\gevcc\, Higgs boson. All four channels show a ratio $\approx\!1$ over the whole NN output range but the {\sl VH}-SS channel has several points with a ratio of $\approx\!0.9$  at the NN output of $\sim\!0.5$;  the most sensitive region of the NN output where the Higgs signal peaks.
%
%
%
%
If the background was mismodeled, either the TRF has incorrectly predicted the QCD background or the NN was at fault. The {\sl VH}-SS and VBF-SS channels share the same TRF and the VBF-SS observed limit agrees with its expected limit. The {\sl VH}-SS and {\sl VH}-SJ channel share the same trained NN and the observed and expected limits in the {\sl VH}-SJ channel agree. Since neither the NN nor TRF showed any evidence of corrupting the background prediction, it suggested the low ratio for {\sl VH}-SS was likely to be a statistical fluctuation rather than evidence of background mismodeling.

\section{Summary}

In summary, a search for the Higgs boson was performed in the all-hadronic final state and set observed (expected) limits of 10.5 (20.0) times the predicted Standard Model cross section at 95\% CL for 120\,\gevcc\, Higgs boson. The measurements presented in this article has shown a factor of two improvement over the previous 2\,\invfb\, result for the all-hadronic Higgs boson search~\cite{PhysRevLett.103.221801} . 
This article extended the 2\,\invfb\, analysis by including the VBF channel, adding an additional algorithm to identify  bottom-quark jets,  adding an artificial neural network to separate signal from background which includes $\langle \phi \rangle$ and $\langle \eta \rangle$ to distinguish gluon jets from quark jets, and by doubling the analyzed data set.  CDF II continues to collect more data and further improvements to the analysis technique will extend the sensitivity of the all-hadronic Higgs boson search.




We thank the Fermilab staff and the technical staffs of the participating institutions for their vital contributions. This work was supported by the U.S. Department of Energy and National Science Foundation; the Italian Istituto Nazionale di Fisica Nucleare; the Ministry of Education, Culture, Sports, Science and Technology of Japan; the Natural Sciences and Engineering Research Council of Canada; the National Science Council of the Republic of China; the Swiss National Science Foundation; the A.P. Sloan Foundation; the Bundesministerium f\"ur Bildung und Forschung, Germany; the Korean World Class University Program, the National Research Foundation of Korea; the Science and Technology Facilities Council and the Royal Society, UK; the Institut National de Physique Nucleaire et Physique des Particules/CNRS; the Russian Foundation for Basic Research; the Ministerio de Ciencia e Innovaci\'{o}n, and Programa Consolider-Ingenio 2010, Spain; the Slovak R\&D Agency; the Academy of Finland; and the Australian Research Council (ARC).

\bibliography{Reference}

\begin{thebibliography}{28}%
\makeatletter
\providecommand \@ifxundefined [1]{%
 \@ifx{#1\undefined}
}%
\providecommand \@ifnum [1]{%
 \ifnum #1\expandafter \@firstoftwo
 \else \expandafter \@secondoftwo
 \fi
}%
\providecommand \@ifx [1]{%
 \ifx #1\expandafter \@firstoftwo
 \else \expandafter \@secondoftwo
 \fi
}%
\providecommand \natexlab [1]{#1}%
\providecommand \enquote  [1]{``#1''}%
\providecommand \bibnamefont  [1]{#1}%
\providecommand \bibfnamefont [1]{#1}%
\providecommand \citenamefont [1]{#1}%
\providecommand \href@noop [0]{\@secondoftwo}%
\providecommand \href [0]{\begingroup \@sanitize@url \@href}%
\providecommand \@href[1]{\@@startlink{#1}\@@href}%
\providecommand \@@href[1]{\endgroup#1\@@endlink}%
\providecommand \@sanitize@url [0]{\catcode `\\12\catcode `\$12\catcode
  `\&12\catcode `\#12\catcode `\^12\catcode `\_12\catcode `\%12\relax}%
\providecommand \@@startlink[1]{}%
\providecommand \@@endlink[0]{}%
\providecommand \url  [0]{\begingroup\@sanitize@url \@url }%
\providecommand \@url [1]{\endgroup\@href {#1}{\urlprefix }}%
\providecommand \urlprefix  [0]{URL }%
\providecommand \Eprint [0]{\href }%
\@ifxundefined \urlstyle {%
  \providecommand \doi  [0]{\begingroup \@sanitize@url \@doi}%
  \providecommand \@doi [1]{\endgroup \@@startlink {\doibase
  #1}doi:\discretionary {}{}{}#1\@@endlink }%
}{%
  \providecommand \doi  [0]{doi:\discretionary{}{}{}\begingroup
  \urlstyle{rm}\Url }%
}%
\providecommand \doibase [0]{http://dx.doi.org/}%
\providecommand \Doi [0]{\begingroup \@sanitize@url \@Doi }%
\providecommand \@Doi  [1]{\endgroup\@@startlink{\doibase#1}\@@Doi}%
\providecommand \@@Doi [1]{#1\@@endlink}%
\providecommand \selectlanguage [0]{\@gobble}%
\providecommand \bibinfo  [0]{\@secondoftwo}%
\providecommand \bibfield  [0]{\@secondoftwo}%
\providecommand \translation [1]{[#1]}%
\providecommand \BibitemOpen [0]{}%
\providecommand \bibitemStop [0]{}%
\providecommand \bibitemNoStop [0]{.\EOS\space}%
\providecommand \EOS [0]{\spacefactor3000\relax}%
\providecommand \BibitemShut  [1]{\csname bibitem#1\endcsname}%
\bibitem [{\citenamefont {Higgs}(1964)}]{Higgs:1964ia}%
  \BibitemOpen
  \bibfield  {author} {\bibinfo {author} {\bibfnamefont {P.~W.}\ \bibnamefont
  {Higgs}},\ }\Doi {10.1016/0031-9163(64)91136-9} {\bibfield  {journal}
  {\bibinfo  {journal} {Phys. Lett.},\ }\textbf {\bibinfo {volume} {12}},\
  \bibinfo {pages} {132} (\bibinfo {year} {1964})}\BibitemShut {NoStop}%
\bibitem [{\citenamefont {Englert}(1964)}]{Englert1964Broken-Symmetry}%
  \BibitemOpen
  \bibfield  {author} {\bibinfo {author} {\bibfnamefont {F.}~\bibnamefont
  {Englert}},\ }\Doi {10.1103/PhysRevLett.13.321} {\bibfield  {journal}
  {\bibinfo  {journal} {Phys. Rev. Lett.},\ }\textbf {\bibinfo {volume} {13}},\
  \bibinfo {pages} {321} (\bibinfo {year} {1964})}\BibitemShut {NoStop}%
\bibitem [{\citenamefont {{The ALEPH, DELPHI, L3 and OPAL Collaborations, and
  the LEP Working Group for Higgs boson searches}}(2003)}]{Barate:2003sz}%
  \BibitemOpen
  \bibfield  {author} {\bibinfo {author} {\bibnamefont {{The ALEPH, DELPHI, L3
  and OPAL Collaborations, and the LEP Working Group for Higgs boson
  searches}}},\ }\Doi {10.1016/S0370-2693(03)00614-2} {\bibfield  {journal}
  {\bibinfo  {journal} {Phys. Lett. B},\ }\textbf {\bibinfo {volume} {565}},\
  \bibinfo {pages} {61} (\bibinfo {year} {2003})}\BibitemShut {NoStop}%
\bibitem [{\citenamefont {Aaltonen}\ \emph
  {et~al.}(2010){\natexlab{a}}\citenamefont {Aaltonen} \emph
  {et~al.}}]{PhysRevLett.104.061802}%
  \BibitemOpen
  \bibfield  {author} {\bibinfo {author} {\bibfnamefont {T.}~\bibnamefont
  {Aaltonen}} \emph {et~al.} (\bibinfo {collaboration} {CDF Collaboration}),\
  }\Doi {10.1103/PhysRevLett.104.061802} {\bibfield  {journal} {\bibinfo
  {journal} {Phys. Rev. Lett.},\ }\textbf {\bibinfo {volume} {104}},\ \bibinfo
  {pages} {061802} (\bibinfo {year} {2010}{\natexlab{a}})}\BibitemShut
  {NoStop}%
\bibitem [{\citenamefont {{CDF Collaboration and D0 Collaboration and Tevatron
  New Physics and Higgs Working Group}}()}]{Collaboration2010Combined-CDF-an}%
  \BibitemOpen
  \bibfield  {author} {\bibinfo {author} {\bibnamefont {{CDF Collaboration and
  D0 Collaboration and Tevatron New Physics and Higgs Working Group}}},\
  }\href@noop {} {\enquote {\bibinfo {title} {Combined {CDF} and {D0} upper
  limits on standard model {H}iggs-boson production with up to 6.7\,fb$^{-1}$
  of data},}\ }\Eprint {http://arxiv.org/abs/hep-ex/1007.4587v1}
  {arXiv:hep-ex/1007.4587v1} \BibitemShut {NoStop}%
\bibitem [{\citenamefont {{ALEPH, CDF, D0, DELPHI, L3, OPAL, SLD, the LEP
  Electroweak Working Group, the Tevatron Electroweak Working Group, and the
  SLD Electroweak and Heavy Flavor Working Groups}}()}]{Alcaraz:2009jr}%
  \BibitemOpen
  \bibfield  {author} {\bibinfo {author} {\bibnamefont {{ALEPH, CDF, D0,
  DELPHI, L3, OPAL, SLD, the LEP Electroweak Working Group, the Tevatron
  Electroweak Working Group, and the SLD Electroweak and Heavy Flavor Working
  Groups}}},\ }\href@noop {} {\enquote {\bibinfo {title} {{Precision
  Electroweak Measurements and Constraints on the Standard Model}},}\ }\Eprint
  {http://arxiv.org/abs/hep-ex/0911.2604} {arXiv:hep-ex/0911.2604} \BibitemShut
  {NoStop}%
\bibitem [{\citenamefont {Djouadi}\ \emph {et~al.}(1998)\citenamefont
  {Djouadi}, \citenamefont {Kalinowski},\ and\ \citenamefont
  {Spira}}]{Djouadi:1997yw}%
  \BibitemOpen
  \bibfield  {author} {\bibinfo {author} {\bibfnamefont {A.}~\bibnamefont
  {Djouadi}}, \bibinfo {author} {\bibfnamefont {J.}~\bibnamefont {Kalinowski}},
  \ and\ \bibinfo {author} {\bibfnamefont {M.}~\bibnamefont {Spira}},\ }\Doi
  {10.1016/S0010-4655(97)00123-9} {\bibfield  {journal} {\bibinfo  {journal}
  {Comput. Phys. Commun.},\ }\textbf {\bibinfo {volume} {108}},\ \bibinfo
  {pages} {56} (\bibinfo {year} {1998})}\BibitemShut {NoStop}%
\bibitem [{\citenamefont {Nakamura}\ and\ \citenamefont
  {Group}(2010)}]{0954-3899-37-7A-075021}%
  \BibitemOpen
  \bibfield  {author} {\bibinfo {author} {\bibfnamefont {K.}~\bibnamefont
  {Nakamura}}\ and\ \bibinfo {author} {\bibfnamefont {P.~D.}\ \bibnamefont
  {Group}},\ }\href {http://stacks.iop.org/0954-3899/37/i=7A/a=075021}
  {\bibfield  {journal} {\bibinfo  {journal} {Journal of Physics G: Nuclear and
  Particle Physics},\ }\textbf {\bibinfo {volume} {37}},\ \bibinfo {pages}
  {075021} (\bibinfo {year} {2010})}\BibitemShut {NoStop}%
\bibitem [{\citenamefont {Aaltonen}\ \emph
  {et~al.}(2009){\natexlab{a}}\citenamefont {Aaltonen} \emph
  {et~al.}}]{Aaltonen2009Search-for-the-}%
  \BibitemOpen
  \bibfield  {author} {\bibinfo {author} {\bibfnamefont {T.}~\bibnamefont
  {Aaltonen}} \emph {et~al.} (\bibinfo {collaboration} {CDF Collaboration}),\
  }\Doi {10.1103/PhysRevD.80.071101} {\bibfield  {journal} {\bibinfo  {journal}
  {Phys. Rev. D},\ }\textbf {\bibinfo {volume} {80}},\ \bibinfo {pages}
  {071101} (\bibinfo {year} {2009}{\natexlab{a}})}\BibitemShut {NoStop}%
\bibitem [{\citenamefont {Aaltonen}\ \emph
  {et~al.}(2009){\natexlab{b}}\citenamefont {Aaltonen} \emph
  {et~al.}}]{PhysRevLett.103.101802}%
  \BibitemOpen
  \bibfield  {author} {\bibinfo {author} {\bibfnamefont {T.}~\bibnamefont
  {Aaltonen}} \emph {et~al.} (\bibinfo {collaboration} {CDF Collaboration}),\
  }\Doi {10.1103/PhysRevLett.103.101802} {\bibfield  {journal} {\bibinfo
  {journal} {Phys. Rev. Lett.},\ }\textbf {\bibinfo {volume} {103}},\ \bibinfo
  {pages} {101802} (\bibinfo {year} {2009}{\natexlab{b}})}\BibitemShut
  {NoStop}%
\bibitem [{\citenamefont {Aaltonen}\ \emph
  {et~al.}(2010){\natexlab{b}}\citenamefont {Aaltonen} \emph
  {et~al.}}]{PhysRevLett.104.141801}%
  \BibitemOpen
  \bibfield  {author} {\bibinfo {author} {\bibfnamefont {T.}~\bibnamefont
  {Aaltonen}} \emph {et~al.} (\bibinfo {collaboration} {CDF Collaboration}),\
  }\Doi {10.1103/PhysRevLett.104.141801} {\bibfield  {journal} {\bibinfo
  {journal} {Phys. Rev. Lett.},\ }\textbf {\bibinfo {volume} {104}},\ \bibinfo
  {pages} {141801} (\bibinfo {year} {2010}{\natexlab{b}})}\BibitemShut
  {NoStop}%
\bibitem [{\citenamefont {Aaltonen}\ \emph
  {et~al.}(2009){\natexlab{c}}\citenamefont {Aaltonen} \emph
  {et~al.}}]{PhysRevLett.103.221801}%
  \BibitemOpen
  \bibfield  {author} {\bibinfo {author} {\bibfnamefont {T.}~\bibnamefont
  {Aaltonen}} \emph {et~al.} (\bibinfo {collaboration} {CDF Collaboration}),\
  }\Doi {10.1103/PhysRevLett.103.221801} {\bibfield  {journal} {\bibinfo
  {journal} {Phys. Rev. Lett.},\ }\textbf {\bibinfo {volume} {103}},\ \bibinfo
  {pages} {221801} (\bibinfo {year} {2009}{\natexlab{c}})}\BibitemShut
  {NoStop}%
\bibitem [{\citenamefont {Acosta}\ \emph
  {et~al.}(2005){\natexlab{a}}\citenamefont {Acosta} \emph
  {et~al.}}]{Acosta:2004yw}%
  \BibitemOpen
  \bibfield  {author} {\bibinfo {author} {\bibfnamefont {D.~E.}\ \bibnamefont
  {Acosta}} \emph {et~al.} (\bibinfo {collaboration} {CDF Collaboration}),\
  }\Doi {10.1103/PhysRevD.71.032001} {\bibfield  {journal} {\bibinfo  {journal}
  {Phys. Rev. D},\ }\textbf {\bibinfo {volume} {71}},\ \bibinfo {pages}
  {032001} (\bibinfo {year} {2005}{\natexlab{a}})}\BibitemShut {NoStop}%
\bibitem [{\citenamefont {Acosta}\ \emph
  {et~al.}(2005){\natexlab{b}}\citenamefont {Acosta} \emph
  {et~al.}}]{Acosta:2004hw}%
  \BibitemOpen
  \bibfield  {author} {\bibinfo {author} {\bibfnamefont {D.~E.}\ \bibnamefont
  {Acosta}} \emph {et~al.} (\bibinfo {collaboration} {CDF Collaboration}),\
  }\Doi {10.1103/PhysRevD.71.052003} {\bibfield  {journal} {\bibinfo  {journal}
  {Phys. Rev. D},\ }\textbf {\bibinfo {volume} {71}},\ \bibinfo {pages}
  {052003} (\bibinfo {year} {2005}{\natexlab{b}})}\BibitemShut {NoStop}%
\bibitem [{\citenamefont {Abulencia}\ \emph {et~al.}(2007)\citenamefont
  {Abulencia} \emph {et~al.}}]{Abulencia:2005ix}%
  \BibitemOpen
  \bibfield  {author} {\bibinfo {author} {\bibfnamefont {A.}~\bibnamefont
  {Abulencia}} \emph {et~al.} (\bibinfo {collaboration} {CDF Collaboration}),\
  }\Doi {10.1088/0954-3899/34/12/001} {\bibfield  {journal} {\bibinfo
  {journal} {J. Phys. G},\ }\textbf {\bibinfo {volume} {34}},\ \bibinfo {pages}
  {2457} (\bibinfo {year} {2007})}\BibitemShut {NoStop}%
\bibitem [{\citenamefont {Abe}\ \emph {et~al.}(1992)\citenamefont {Abe} \emph
  {et~al.}}]{Abe:1991ui}%
  \BibitemOpen
  \bibfield  {author} {\bibinfo {author} {\bibfnamefont {F.}~\bibnamefont
  {Abe}} \emph {et~al.} (\bibinfo {collaboration} {CDF Collaboration}),\ }\Doi
  {10.1103/PhysRevD.45.1448} {\bibfield  {journal} {\bibinfo  {journal} {Phys.
  Rev. D},\ }\textbf {\bibinfo {volume} {45}},\ \bibinfo {pages} {1448}
  (\bibinfo {year} {1992})}\BibitemShut {NoStop}%
\bibitem [{\citenamefont {Bhatti}\ \emph {et~al.}(2006)\citenamefont {Bhatti}
  \emph {et~al.}}]{Bhatti:2005ai}%
  \BibitemOpen
  \bibfield  {author} {\bibinfo {author} {\bibfnamefont {A.}~\bibnamefont
  {Bhatti}} \emph {et~al.} (\bibinfo {collaboration} {CDF Collaboration}),\
  }\href@noop {} {\bibfield  {journal} {\bibinfo  {journal} {Nucl. Instrum.
  Methods},\ }\textbf {\bibinfo {volume} {A566}},\ \bibinfo {pages} {375}
  (\bibinfo {year} {2006})}\BibitemShut {NoStop}%
\bibitem [{Note1()}]{Note1}%
  \BibitemOpen
  \bibinfo {note} {Missing transverse energy significance is defined as the
  ratio of the total missing transverse energy to the square root of the total
  transverse energy}\BibitemShut {NoStop}%
\bibitem [{\citenamefont {Abulencia}\ \emph {et~al.}(2006)\citenamefont
  {Abulencia} \emph {et~al.}}]{Abulencia:2006kv}%
  \BibitemOpen
  \bibfield  {author} {\bibinfo {author} {\bibfnamefont {A.}~\bibnamefont
  {Abulencia}} \emph {et~al.} (\bibinfo {collaboration} {CDF Collaboration}),\
  }\Doi {10.1103/PhysRevD.74.072006} {\bibfield  {journal} {\bibinfo  {journal}
  {Phys. Rev. D},\ }\textbf {\bibinfo {volume} {74}},\ \bibinfo {pages}
  {072006} (\bibinfo {year} {2006})}\BibitemShut {NoStop}%
\bibitem [{\citenamefont {Sj\"ostrand}\ \emph {et~al.}(2001)\citenamefont
  {Sj\"ostrand} \emph {et~al.}}]{Sjostrand:2000wi}%
  \BibitemOpen
  \bibfield  {author} {\bibinfo {author} {\bibfnamefont {T.}~\bibnamefont
  {Sj\"ostrand}} \emph {et~al.},\ }\Doi {10.1016/S0010-4655(00)00236-8}
  {\bibfield  {journal} {\bibinfo  {journal} {Comput. Phys. Commun.},\ }\textbf
  {\bibinfo {volume} {135}},\ \bibinfo {pages} {238} (\bibinfo {year}
  {2001})}\BibitemShut {NoStop}%
\bibitem [{\citenamefont {Brun}\ \emph {et~al.}(1987)\citenamefont {Brun} \emph
  {et~al.}}]{GEANT}%
  \BibitemOpen
  \bibfield  {author} {\bibinfo {author} {\bibfnamefont {R.}~\bibnamefont
  {Brun}} \emph {et~al.},\ }\href@noop {} {\bibfield  {journal} {\bibinfo
  {journal} {{CERN}-{DD}-{EE}-84-01}} (\bibinfo {year} {1987})}\BibitemShut
  {NoStop}%
\bibitem [{\citenamefont {{Gerchtein}}\ and\ \citenamefont
  {{Paulini}}()}]{2003physics...6031G}%
  \BibitemOpen
  \bibfield  {author} {\bibinfo {author} {\bibfnamefont {E.}~\bibnamefont
  {{Gerchtein}}}\ and\ \bibinfo {author} {\bibfnamefont {M.}~\bibnamefont
  {{Paulini}}},\ }\href@noop {} {}\Eprint {http://arxiv.org/abs/hep-ex/0306031}
  {arXiv:hep-ex/0306031} \BibitemShut {NoStop}%
\bibitem [{\citenamefont {Aaltonen}\ \emph
  {et~al.}(2010){\natexlab{c}}\citenamefont {Aaltonen} \emph
  {et~al.}}]{Aaltonen2010Measurement-of-}%
  \BibitemOpen
  \bibfield  {author} {\bibinfo {author} {\bibfnamefont {T.}~\bibnamefont
  {Aaltonen}} \emph {et~al.} (\bibinfo {collaboration} {CDF Collaboration}),\
  }\Doi {10.1103/PhysRevD.81.052011} {\bibfield  {journal} {\bibinfo  {journal}
  {Phys. Rev. D},\ }\textbf {\bibinfo {volume} {81}},\ \bibinfo {pages}
  {052011} (\bibinfo {year} {2010}{\natexlab{c}})}\BibitemShut {NoStop}%
\bibitem [{\citenamefont {Aaltonen}\ \emph
  {et~al.}(2010){\natexlab{d}}\citenamefont {Aaltonen} \emph
  {et~al.}}]{Collaboration2010First-Measureme}%
  \BibitemOpen
  \bibfield  {author} {\bibinfo {author} {\bibfnamefont {T.}~\bibnamefont
  {Aaltonen}} \emph {et~al.} (\bibinfo {collaboration} {CDF Collaboration}),\
  }\Doi {10.1103/PhysRevLett.105.012001} {\bibfield  {journal} {\bibinfo
  {journal} {Phys. Rev. Lett.},\ }\textbf {\bibinfo {volume} {105}},\ \bibinfo
  {pages} {012001} (\bibinfo {year} {2010}{\natexlab{d}})}\BibitemShut
  {NoStop}%
\bibitem [{\citenamefont {Hoecker}\ \emph {et~al.}()\citenamefont {Hoecker}
  \emph {et~al.}}]{HoeckerTMVA---Toolkit-}%
  \BibitemOpen
  \bibfield  {author} {\bibinfo {author} {\bibfnamefont {A.}~\bibnamefont
  {Hoecker}} \emph {et~al.},\ }\href@noop {} {}\Eprint
  {http://arxiv.org/abs/hep-ex/0703039v5} {arXiv:hep-ex/0703039v5} \BibitemShut
  {NoStop}%
\bibitem [{Note2()}]{Note2}%
  \BibitemOpen
  \bibinfo {note} {$\protect \qopname \relax o{cos}\theta ^{*}_ {q_{1}}$ is the
  cosine helicity angle of $q_1$. The $q_1$ helicity angle, $ \theta
  ^{*}_{q_{1}}$, is defined to be the angle between the momentum of $q_1$ in
  the $q_1-q_2$ rest frame and the total momentum of $q_1-q_2$ in the lab
  frame.}\BibitemShut {Stop}%
\bibitem [{Note3()}]{Note3}%
  \BibitemOpen
  \bibinfo {note} {$\protect \qopname \relax o{cos}{\theta _{3}}$ is defined in
  a three jet rest frame as the cosine of the leading jet scattering angle. We
  reduce from four jets to three jets by combining the two jets with the lowest
  dijet mass. Thus $\protect \qopname \relax o{cos}{\theta _{3}} = \protect
  \frac {\protect \mathaccentV {vec}17E{P_{AV}} \cdot \protect \mathaccentV
  {vec}17E{P_{3}}}{|\protect \mathaccentV {vec}17E{P_{AV}}|\protect
  \mathaccentV {vec}17E{P_{3}}|}$, where $\protect \mathaccentV
  {vec}17E{P_{3}}$ is the third jet and $\protect \mathaccentV
  {vec}17E{P_{AV}}$ is the vector sum of the three jets in the lab frame~\cite
  {Geer:1995mp}.}\BibitemShut {Stop}%
\bibitem [{\citenamefont {Geer}\ and\ \citenamefont
  {Asakawa}(1996)}]{Geer:1995mp}%
  \BibitemOpen
  \bibfield  {author} {\bibinfo {author} {\bibfnamefont {S.}~\bibnamefont
  {Geer}}\ and\ \bibinfo {author} {\bibfnamefont {T.}~\bibnamefont {Asakawa}},\
  }\Doi {10.1103/PhysRevD.53.4793} {\bibfield  {journal} {\bibinfo  {journal}
  {Phys. Rev.},\ }\textbf {\bibinfo {volume} {D53}},\ \bibinfo {pages} {4793}
  (\bibinfo {year} {1996})}\BibitemShut {NoStop}%
\end{thebibliography}%

\end{document}